\documentclass{article} 

\usepackage[utf8]{inputenc} 
\usepackage[T1]{fontenc}    
\usepackage{hyperref}       
\usepackage{url}            
\usepackage{booktabs}       
 \usepackage{amsfonts}       
\usepackage{nicefrac}       
\usepackage{microtype}      
\usepackage{lipsum}
\usepackage[colorinlistoftodos]{todonotes}

\usepackage{afterpage} 
\usepackage{etex}
\usepackage{color}
\usepackage[stable]{footmisc} 
\usepackage[section]{placeins} 
							
\usepackage{graphicx} 
\usepackage{multirow} 
\usepackage{longtable}

\usepackage{amsmath}
\usepackage{amssymb}
\usepackage{amstext}
\usepackage{amsthm}
\usepackage{enumerate}
\usepackage{bbold}
 \usepackage{extarrows}
\usepackage{esvect}

\usepackage{relsize}
\usepackage{array}
\usepackage{dsfont}

\usepackage{pstricks-add}
\usepackage{changes}
\usepackage{caption}
\usepackage{subcaption}	
\usepackage{float}

\usepackage{parskip}
\usepackage{stmaryrd}
\usepackage{MnSymbol}

\usepackage{cancel}
\usepackage{framed} 
\usepackage{fancyhdr}
\usepackage{arxiv}

\newtheorem*{rem*}{Remark}

\definechangesauthor[name=Bernd, color=MyGreen]{Bernd}
\definechangesauthor[name=Barbara, color=ElectricPurple]{Barbara}

\newcommand{\bra}[1]{\left < #1 \right |}
\newcommand{\ket}[1]{\left | #1 \right>}
\newcommand{\ScalProd}[2]{\left< #1 \,|\, #2 \right>}
\newcommand{\N}{\mathbb{N}} 
\newcommand{\R}{\mathbb{R}} 
\newcommand{\C}{\mathbb{C}} 
\renewcommand{\d}{\text{ d}} %
\newcommand{\s}{\mathcal{S}} %
\newcommand{\Edge}{\mathcal{E}} %
\newcommand{\LL}{\mathcal{L}} %
\newcommand{\Id}{\mathbb{1}} 
\newcommand{\G}{\Gamma} %
\newcommand{\g}{\gamma} %
\newcommand{\e}{\mathrm{e}} %

\newcommand{\bp}{\boldsymbol{p}}
\newcommand{\bq}{\boldsymbol{q}}
\newcommand{\bh}{\boldsymbol{h}}
\newcommand{\brho}{\boldsymbol{\rho}}
\newcommand{\be}{\boldsymbol{e}}

\newcommand{\bzero}{\boldsymbol{0}}
\newcommand{\bTheta}{\boldsymbol{\Theta}}

\newcommand{\Unravel}{\mathcal{U}}

\newcommand{\Tr}{\text{Tr}}
\newcommand{\Prob}{\mathcal{P}}

\newcommand{\Span}{\text{span }}
\newcommand{\Kern}{\text{kern }}

\renewcommand{\Im}{\text{Im }}
\newcommand{\Image}{\text{image }}
\newcommand{\InTree}{\mathcal{T}}

\newcount\colveccount
\newcommand*\colvec[1]{
        \global\colveccount#1
        \begin{pmatrix}
        \colvecnext
}
\def\colvecnext#1{
        #1
        \global\advance\colveccount-1
        \ifnum\colveccount>0
                \\
                \expandafter\colvecnext
        \else
                \end{pmatrix}
        \fi
}

\definecolor{MyYellow}{rgb}{1.0, 0.75, 0.0}
\definecolor{MyGreen}{rgb}{0.0, 0.5, 0.0}
\definecolor{CrimsonGlory}{rgb}{0.75, 0.0, 0.2}
\definecolor{DarkRed}{rgb}{0.55, 0.0, 0.0}
\definecolor{ElectricPurple}{rgb}{0.75, 0.0, 1.0}

\definecolor{TUDa_0c}{RGB}{137, 137, 137}

\definecolor{TUDa_1a}{RGB}{93, 133, 195} 
\definecolor{TUDa_1b}{RGB}{0, 90, 169} 
\definecolor{TUDa_1c}{RGB}{0, 78, 138} 
\definecolor{TUDa_1d}{RGB}{36, 53, 114} 

\definecolor{TUDa_2a}{RGB}{0, 156, 218}
\definecolor{TUDa_2b}{RGB}{0, 131, 204}
\definecolor{TUDa_2c}{RGB}{0, 104, 157}
\definecolor{TUDa_2d}{RGB}{0, 78, 115}

\definecolor{TUDa_4b}{RGB}{153, 192, 0} 
\definecolor{TUDa_4d}{RGB}{106, 139, 55} 

\definecolor{TUDa_9a}{RGB}{233, 80, 62} 
\definecolor{TUDa_9b}{RGB}{230, 0, 26} 
\definecolor{TUDa_9c}{RGB}{185, 15, 34} 

\definecolor{TUDa_3a}{RGB}{80, 182, 140} 

\definecolor{MyGreen}{rgb}{0.0, 0.5, 0.0}
\definecolor{MyPurple}{RGB}{141, 0, 255} 
\definecolor{MyCyan}{rgb}{0, 0.67, 0.67} 

\definecolor{TUDa_11a}{RGB}{128, 69, 151} 
\definecolor{TUDa_11b}{RGB}{114, 16, 133} 
\definecolor{TUDa_11c}{RGB}{97, 28, 115}  
\definecolor{TUDa_11d}{RGB}{76, 34, 106}  

\definecolor{TUDa_10a}{RGB}{201, 48, 142} 
\definecolor{TUDa_10b}{RGB}{166, 0, 132} 
\definecolor{TUDa_10c}{RGB}{149, 17, 105}  
\definecolor{TUDa_10d}{RGB}{115, 32, 84}  

\begin{document}

\title{  
 Explicit expressions for stationary states of the Lindblad equation for a finite state space }
\author{
Bernd Fernengel  \\
Institute of Condensed Matter Physics \\
Technical University of Darmstadt \\
Hochschulst. 6, 64289 Darmstadt, Germany \\
\texttt{bernd@pkm.tu-darmstadt.de} \\
  \And
Barbara Drossel  \\
Institute of Condensed Matter Physics \\
Technical University of Darmstadt \\
Hochschulst. 6, 64289 Darmstadt, Germany \\ 
\texttt{drossel@pkm.tu-darmstadt.de} \\
} 

\maketitle

\begin{abstract}
The Lindblad equation describes the time evolution of a density matrix
of a quantum mechanical system. Stationary solutions are obtained by
time-averaging the solution, which will in general depend on the initial
state. We provide an analytical expression for the steady states of the
Lindblad equation using the quantum jump unraveling, a version of an
ergodic theorem, and the stationary probabilities of the corresponding
discret-time Markov chains. Our result is valid when the
number of states appearing the in quantum trajectory is finite. The classical case of a
Markov jump-process is recovered as a special case, and differences
between the two are discussed.
\end{abstract}

\section{Introduction} 

The Lindblad equation is a quantum master equation describing the time evolution of a density matrix of an open quantum system. It is given by

\begin{equation}\label{Lindbladian}
\begin{aligned}
\partial_t \brho(t) 
&=
\LL(\brho) 
=
\underbrace{
-i [H, \brho(t)]
}_{
\text{von Neumann term}
}
+
\underbrace{
\sum\limits_{k \in I} \g_k \left( V_k\, \brho(t)\, V_k^\dagger - \frac{1}{2} \left\{V_k^\dagger V_k, \brho(t) \right\} \right)
}_{
\text{dissipator term}
} \\
&=
-i \left(H_c \, \brho(t) - \brho(t) \, H_c  \right) + \sum\limits_{k \in I} \g_k \left( V_k \, \brho(t) \, V_k^\dagger  \right)\, , \\
\brho(t=0) &= \brho_0, 
\end{aligned}
\end{equation}

where $[\cdot,\cdot]$ and $\{\cdot,\cdot\}$ denote the commutator and anticommutator, respectively, and the so-called \textit{conditional Hamiltonian} $H_c$ is defined as 

\begin{align}\label{Def_H_c}
H_c 
:=
H - \frac{i}{2}  \sum\limits_{k \in I} \g_k \left( V_k^\dagger \, V_k \right) 
=:
H - \frac{i}{2} \, \Lambda.  
\end{align}

We have set $\hbar = 1$. An open quantum system is a quantum mechanical system in contact with an external environment. 
The Lindbladian $\LL(\brho)$ of equation \eqref{Lindbladian} is the most general form of an (infinitesimal) generator that ensures that the solution $\brho(t)$ remains a density matrix, in particular that is positive semi-definite (with respect to the complex numbers) and has trace one. This theorem has first been proven by Gorini, Kossakowski, and Sudarshan for the finite dimensional case, with Lindblad generalizing the statement to bounded generators \cite{breuer2002theory}. At that time, generators of the form $\LL(\brho)$ have already appeared multiple times in the physical literature \cite{breuer2002theory}. 
In addition to these previous theorems, which are based on quantum Markov semi-groups, there are two types of microscopic derivations of the Lindblad equation, namely the weak-coupling limit and the low-density limit. The basis for both derivations is the ad-hoc assumption that the combined system of system and bath can be written initially as a product state. From this initial state, one computes the dynamics and traces over the environmental degrees of freedom to arrive at the reduced density matrix of the system. 

When conducting a numerical simulation of a specific model, one often relies on \textit{unravellings} of the Lindblad equation, which are ensembles of stochastic trajectories whose average yields the solution of the Lindblad equation \cite{breuer2002theory}. While different unravellings may yield qualitative different types of dynamics, all measurable quantities depend only on the solution $\brho(t)$, while additional properties depending on 
either a certain type of unravelling or a specific quantum trajectory
are not accessible to experimental observation. 

The Lindblad equation contains the von Neumann equation, describing a closed quantum mechanical system, as a special case. The solution von the von Neumann equation is given by $\e^{-i \, H \, t} \,  \brho_0 \, \e^{i \, H \, t} $, with the self-adjoint Hamiltonian $H^\dagger = H$. Since $H_c$ is not self-adjoint, the (conditional) time evolution operator $\e^{-i\,H_c \, t}$ is not unitary.

The additional terms can be interpreted as modeling the influence of the environment, in particular by inducing transitions between states of the system. The so-called \textit{Lindblad operators} $\{V_k \,|\, k \in I \}$ can be interpreted as causing transitions $\psi \to \frac{V_k \, \psi}{\| V_k \, \psi \|}$, where $\emptyset \neq I \subset \N$ is a finite index set and $\g_k > \, 0$ are positive transition rates. We will restrict ourselves to the finite-dimensional case, that is $\brho, V_k, H \in \C^{N \times N}$, with $N \in \N, N\geq \, 2$.

One important aspect in the study of quantum master equation is its long-term behaviour $\lim\limits_{t \to \infty} \brho(t)$ and its steady states $\brho^\infty(\brho_0)$ ( with $\brho^\infty$ defined by $\LL(\brho^\infty) = 0$), as this determines the state where a quantum mechanical system will eventually be. When certain algebraic conditions on the Lindblad operators are satisfied, the steady state is unique and asymptotically stable. The most prominent theorem about such conditions was given by Spohn \cite{spohn1977algebraic}. It requires that the set of Lindblad operators form a basis, i.e. that the environment couples to all degrees of freedom, which applies only to a limited set of physical systems and cannot easily be generalized. 

Most papers studying the stationary states of the Lindblad equation require additional properties of the Lindblad operators. Bua and Prosen propose a method which assumes that the underlying symmetry of the system is already known \cite{buvca2012note}. While being numerically cheap, finding all the symmetries of an open system can be highly non-trivial, especially when both "strong" and "weak" symmetry is involved (see \cite{thingna2021degenerated} for a summary). 
Another approach was made by Trushechkina, where special kinds of Lindblad operators were considered, which allow a "backwards" transition for every "forward" transition (see \cite{trushechkin2018finding}. For the dimension $N=2$ a full discussion can be found in \cite{andrianov2022franke}. 

The general nature of the Lindbladian in equation \eqref{Lindbladian} makes it difficult to compute 
analytical expressions for all possible steady states. In this paper, we aim at 
statements about the stationary states, without making additional assumptions on the Lindblad operators.
We will address the nature and number of these stationary states, how they can be determined, and how they are related to the long-term behaviour that results from a given initial state.

The standard way to compute the stationary state from a given initial state $\brho_0$ would be to determine the solution $t \mapsto \brho(t)$ of the initial value problem of equation \eqref{Lindbladian} for all positive times $t \geq 0$ and to evaluate the time average $\langle \brho_{\brho_0}(t) \rangle_{t\geq 0} = \lim\limits_{T \to \infty}
\frac{1}{T}
\int_0^T 
\brho_{\brho_0}(t) \, \d \,t$. 
We propose a different method: We  express the density matrix $\brho(t)$ in terms of the average of an ensemble of stochastic quantum trajectories. Then we make use of the fact that the limit of the time average of a single quantum trajectory exists and that it may be interchanged with the ensemble limit \cite{kuemmerer2004pathwise}. We will show that both the time average and the ensemble average of a quantum trajectory can be explicitly evaluated with the theory of discrete-time Markov chains when certain conditions (in particular a finite state space) are satisfied.

\section{The concept of an unravelling of the Lindblad equation}\label{secunravel}

Unravellings are ensembles of stochastic quantum trajectories that are equivalent to the Lindblad equation in the sense that the state of the system $ \brho(t)$ at a time $t\geq 0$ is given by the average over all possible trajectories,

\begin{equation} \label{rho_unravel}
\begin{aligned}
\brho(t)
&=
\int_\Unravel \Theta(t, \omega) \d \, \Prob(\omega).
\end{aligned}
\end{equation}

\begin{figure}[H]
\centering
\includegraphics[width = 0.5\columnwidth]{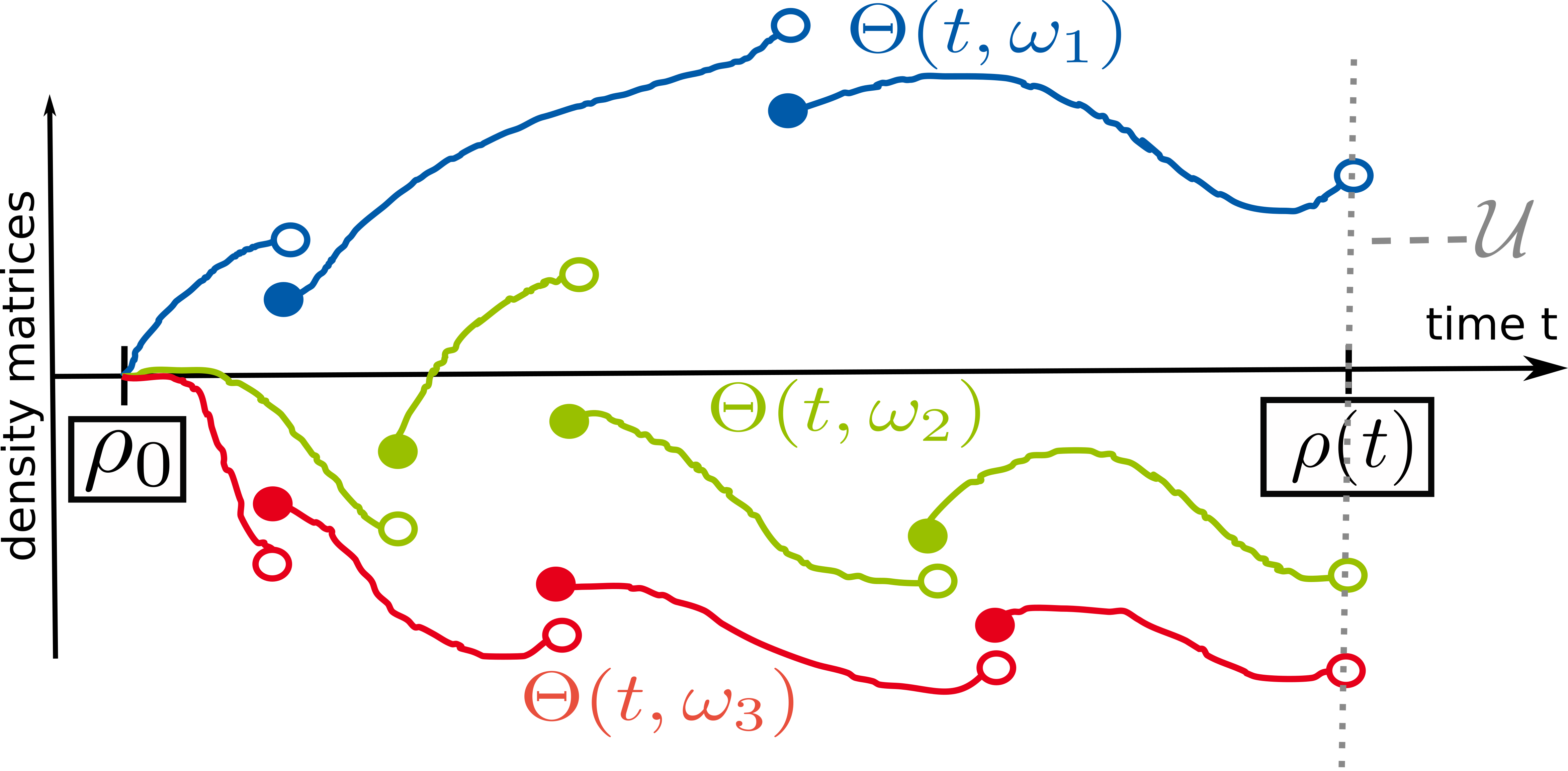}
\caption{Illustration of an ensemble of quantum trajectories, originating from the quantum jump unravelling, which is a special type of piecewise deterministic processes interrupted by quantum jumps. The average of the trajectories
at a given time $t\geq 0$ yields the density matrix $\brho(t)$. }
\label{Illustrating_QuantumTrajectories}
\end{figure}

The parameter $\omega\in\Unravel$ labels the quantum trajectories $\R_{\geq \, 0} \ni \, t   \mapsto \Theta(t, \omega) \in \{\text{density matrices}\}$ (see figure \eqref{Illustrating_QuantumTrajectories} for all illustration).

In particular, the expectation values of any quantum mechanical observable $A$, given the system is in the state $\brho$, can be computed from these unravelings, according to

\begin{equation}
\begin{aligned}
E_{\brho}[A]
&=
\Tr[\brho \, A]
=
\int_\Unravel \Tr [\Theta(\omega)\, A]  \d \, \Prob(\omega). 
\end{aligned}
\end{equation}

Unravellings provide different perspectives on the subject of time evolution of open quantum systems and can yield new insights since they are tackled by a different set of mathematical tools: Instead of linear differential equations one has the theory of Markov chains, piecewise deterministic processes, and stochastic differential equations at hand \cite{breuer2002theory}. 

Moreover, they are often superior when it comes to  numerical simulations of specific models as  a Ket-state unravelling (see section \ref{KetStateUnravellings}) requires less memory storage  (linear with the dimension of the system, instead of quadratic \cite{lidar2019lecture}).

While unravellings are usually done in terms of ket states, we want to focus on the so called \textit{mixed-state unravellings}, which assume values in the set of density matrices \cite{breuer2002theory}. 

Among the different possible types of unravellings, we will use the \textit{quantum-jump unravelling}.  For a given Hamiltonian $H$ and a given set of Lindblad operators $\{V_k \,:\, k \in I\}$, the quantum jump unravelling is a piecewise deterministic process (see \cite{breuer2002theory, lidar2019lecture}) where the continuous,  deterministic time evolution is interrupted by discontinuous quantum jumps at times $t_n$, where a Lindblad operator $V_{\pi_n}$ is applied. 

So a quantum trajectory is completely determined by the sequence of times $(t_n)_{n \in \N} \subset \R^\N$ (which record when a quantum jump takes place) and the sequence $(\pi_n)_{n \in \N} \in $ of indices of Lindblad operators (which records which operator has been applied at time $t_n$). 
We define $\Unravel$ to be the set of all possible quantum trajectories, $\tau_n := t_{n+1} - t_{n}$ to be the 
length of the time interval 
between two jumps and $\Theta_n := \Theta(t_{n})$ the state of the unravelling at time $t_{n}$.

Note that we use with $\Theta(t, \omega)$ a different symbol for a state in the quantum trajectory $\omega \in \Unravel$, to distinguish it from the solution $\brho(t)$ of the Lindblad equation \eqref{Lindbladian} at time $t_n$, which is the average over all quantum trajectories, $\brho(t) = \langle \Theta(\omega, t) \rangle_{\omega \in \Unravel}$.

\subsection{Definitions and preparatory considerations}

\subsubsection{(Pseudo-) Time-evolution operators and quantum jump operators}

To avoid unnecessarily complicated notation, we denote by $U_t$ the (non-unitary) time evolution according to the conditional Hamiltonian $H_c$, and by $J_k$ the quantum jump operator, irrespective of the nature of the quantum state, be it a density matrix or a ket state. Their action on these two types of quantum states is thus given by

\begin{center}
\begin{minipage}[t]{0.3\textwidth}
\begin{equation*}
\begin{aligned}
U_t, J_k &: \C^{N \times N} \to \C^{N \times N} \\
U_t(\Theta) &= \e^{-i \, H_c \, t} \, \Theta \, \e^{i \, H_c^\dagger \, t} \\
J_k (\Theta) &= V_k \, \Theta \, V_k^\dagger
\end{aligned}
\end{equation*}
\end{minipage}\begin{minipage}[t]{0.3\textwidth}
\begin{equation}
\begin{aligned}
U_t, J_k &: \C^{N } \to \C^{N } \\
U_t(\psi) &= \e^{-i \, H_c \, t} \, \psi \\
J_k (\psi) &= V_k \, \psi
\end{aligned}
\end{equation}
\end{minipage}
\end{center}

During the time interval $[t_n, t_{n+1})$ between two quantum jumps there is a deterministic time evolution $\Theta(t) = \frac{U_{t-t_n}\, (\Theta_n)}{\Tr[\dots]}$, and directly after a quantum jump induced by the Lindblad operator $V_k$ the state of the quantum trajectory is given by $\Theta(t_n) = \frac{J_k(\Theta)}{\Tr[\dots]}$ $\Bigl($ we write $\frac{A}{\Tr[\dots]}$ instead of $\frac{A}{\Tr[A]}\Bigr)$.

\subsubsection{The distribution of the waiting time}
Given that the quantum trajectory is in state $\Theta$, the waiting time $t$ for the next jump is distributed according to 

\begin{equation}\label{WaitingTimeDistribution}
\begin{aligned}
f(t \, | \, \Theta) 
&:=
- \frac{d}{dt} \Tr \left[
\underbrace{
U_t(\Theta)
}_{
\e^{-i \, H_c \, t} \, \Theta \, \e^{i \, H_c^\dagger \, t} 
}\right]
=
- \Tr \left[ \underbrace{(-i\, H_c)}_{-i \, H - \frac{1}{2} \, \Lambda} \,U_t(\Theta) + U_t(\Theta) \,  \underbrace{(i\, H_c^\dagger)}_{i \, H - \frac{1}{2} \, \Lambda} \right] = \\
%
&=
\underbrace{i \, \Tr \left[H \, U_t(\Theta) \right] -  i \, \Tr \left[ U_t(\Theta) \, H \right]}_{0} 
+ \underbrace{\frac{1}{2} \Tr \left[\Lambda \, U_t(\Theta) \right] + \frac{1}{2} \Tr \left[ U_t(\Theta) \, \Lambda  \right]}_{\Tr \left[\Lambda \, U_t(\Theta) \right]} = \\
%
&=
\Tr[\Lambda \, U_t(\Theta)] 
\xlongequal{\Lambda = \sum\limits_{k \in I} \g_k \, V_k^\dagger \, V_k}
=
\sum\limits_{k \in I} 
\g_k \,\Tr \left[
\underbrace{
V_k \, \e^{-i \, H_c \, t} \, \Theta \, \e^{i \, H_c^\dagger \, t} \, V_k^\dagger
}_{
J_k \circ U_t(\Theta)
}\right] = \\
&= 
\sum\limits_{k \in I}  
\underbrace{
\g_k \, \Tr \left[
J_k \circ U_t(\Theta)
\right]
}_{
f^{(k)}(t \, | \, \Theta) 
}
=: \sum\limits_{k \in I} f^{(k)}(t \, | \, \Theta) 
\end{aligned}
\end{equation}

with

\begin{align}\label{Def_f_k}
f^{(k)}(t \, | \, \Theta) 
:=
\g_k \, J_k \circ U_t(\Theta) 
=
\g_k \, V_k \, \e^{-i \, H_c \, t} \, \Theta \, \e^{i \, H_c^\dagger \, t} \, V_k^\dagger \;\; \text{ (see \cite{breuer2002theory})}. 
\end{align}

Note that $ f(\cdot \, | \, \Theta)  $ need not be  a probability distribution since $\int_{\R_{\geq \, 0}} f(t \, | \, \Theta) \, \d t  < 1$ is possible. In this case, we call $\Theta$ a \textit{possible trapping state}, which will be discussed below.

\subsubsection{The time-weighted average state}
For a finite waiting time $\tau<\infty$ we define the time-weighted average state between two jumps as 

\begin{equation}\label{Def_I_ThetaOfTau}
\begin{aligned}
I_\Theta(\tau) 
&:= 
\int_0^\tau \, \frac{ U_t(\Theta) }{\Tr[U_t(\Theta)]} \, d \, t
=
\tau \cdot \langle \Theta(t)\rangle_{t \in [0, \tau]}, 
\end{aligned}
\end{equation}
which is the waiting time $\tau$ times the average state between two quantum jumps.


\subsection{The algorithm for an unravelling}\label{AlgorithmForunravelling}

The algorithm for obtaining an unravelling is the following (see \cite{lidar2019lecture} or\cite{breuer2002theory} for a detailed proof): 

\begin{itemize}
\item[1)] Choose an initial state $\brho_0 $ and set $\Theta_0:= \brho_0 = \brho(t_0)$. 
\item[2)] 
Given that the system at time $t_n$ is in state $\Theta_n$, we calculate the waiting time $\tau_n$ to the next jump according to the following considerations: 
The probability for $\tau_n$ to lie in the interval $[0,T)$ is given by

\begin{equation} \label{Prob_WaitingTime}
\begin{aligned}
\Prob \, \bigl( \tau_n \in [0, T) \,|\, \Theta_n \bigr) =
\int_0^T  
\underbrace{\hspace*{5mm} f(t \,|\, \Theta_n  )\hspace*{5mm}}_{ - \frac{d}{dt} \Tr \left[ U_t({\Theta_n}) \right] }
\d t
= 
1 - \Tr \left[ U_T({\Theta_n}) \right] .  
\end{aligned}
\end{equation}

One way of determining the waiting time is the so-called \textit{inversion method} (see: \cite{breuer2002theory}): We draw a random number $\eta_n$ from the uniform distribution of the interval $[0,1]$ and set 

\begin{equation}
\begin{aligned}
\tau_n
:= 
\begin{cases}
f^{-1}(\eta_n \,|\, \Theta_n) &, \text{  if } \eta_n \in \Image\bigl(f(\cdot \,|\, \Theta_n)\bigr) \\
\infty&, \text{ otherwise}
\end{cases}
\end{aligned}
\end{equation}

An illustration of this procedure is given in figure \ref{Example_ChoosingWaitingTime}.

\begin{figure}[H]
\centering
\includegraphics[width = 0.7\columnwidth]{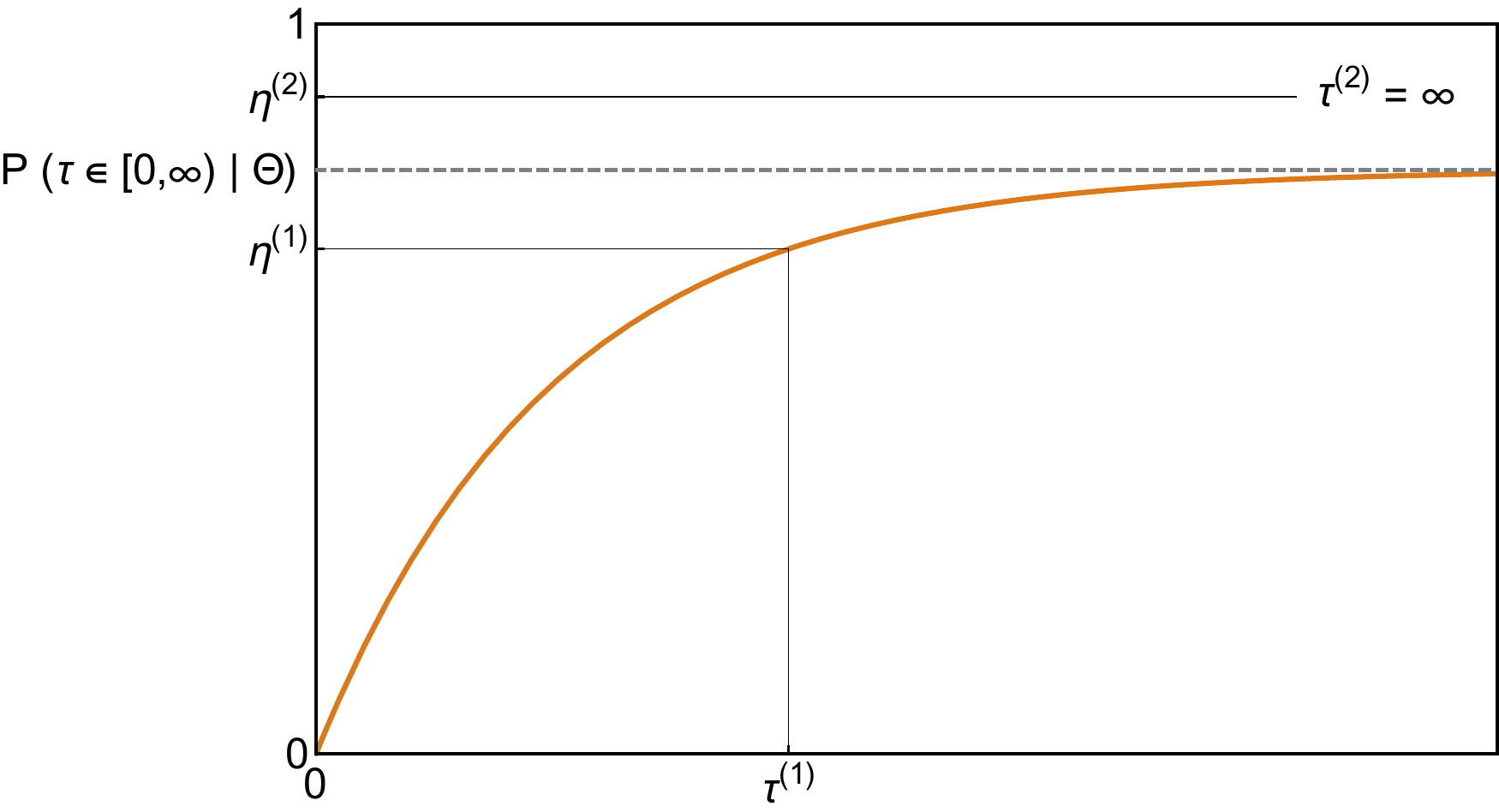}
\caption{ Determining the waiting time $\tau \in \R_{\geq \, 0} \cup \{\infty\}$, given a state $\Theta$: Choose a random variable $\eta$ uniformly distributed  over the interval $[0,1]$ and determine the positive number $\tau$  that satisfies $f(\tau \, | \, \Theta)  = \eta$. If no such number exists, set $\tau = \infty$. The two cases are indicated in the figure: When the random number is $\eta^{(1)}$, the corresponding waiting time is $\tau^{(1)}$; when it is $\eta^{(2)}$ the waiting time is infinite, $\tau^{(2)}=\infty$. 
}
\label{Example_ChoosingWaitingTime}
\end{figure}

Now we have to distinguish two cases: 

\begin{itemize}
\item[i)]
When $\lim\limits_{t \to \infty} \Tr[U_t(\Theta_n)]=0$, then the quantum jump will occur at a finite time. 
\item[ii)]
When $\lim\limits_{t \to \infty} \Tr[U_t(\Theta_n)] \in (0,1]$, then there is a nonzero chance that the process no longer jumps, i.e. the waiting time is infinite. In this case, the state of the quantum trajectory after the time $t_n$ is given by $\frac{U_{t - t_n}(\Theta_n)}{\Tr[\dots]}$, and we call $\Theta_n$ a \textit{possible trapping state}.
\end{itemize}

\item[3)] 
If the waiting time  is finite $(\tau_n < \infty)$  , choose the operator $V_{\pi_n}$ that is applied at time $t_n$ according to

\begin{equation} \label{ProbForNextJump}
\begin{aligned}
\Prob \, 
\bigl(
\pi_n =k \,|\, \Theta_n,  \tau=\tau_n
\bigr) 
=
\frac{
f^{(k)}(\tau_n \, | \, \Theta_n )
}{
\sum\limits_{{j}\in J }f^{(j)}(\tau_n \, | \, \Theta_n )
} , 
\end{aligned}
\end{equation}

with the $f^{(k)}(\cdot \, |\, \Theta )$ defined in equation \eqref{Def_f_k}. The next state is then given by $\Theta_{n+1} = \frac{J_k\, \bigl( U_{\tau_n}(\Theta_n)\bigr)}{\Tr[\dots]}$. 

Then for a fixed unravelling $ \omega \in \Unravel$,  for positive times $t_n \in \R_{\geq 0}$, and (possibly infinite) times $t_{n+1} \in \R_{> \, 0} \cup \{ \infty\}$,  we have

\begin{equation} \label{StatesInAQuantumTrajectory}
\begin{aligned}
\text{  for } {\color{black}t} \in [t_{n}, t_{n+1}):
\Theta_{\brho_0}({\color{black}t}, \omega) 
&= 
\frac{U_{{\color{black}t}-t_{n}} \, (\Theta_n)}{\Tr[\dots]}
= 
\frac{U_{{\color{black}t}-t_{n}} \circ J_{\pi_{n}} \, \circ U_{\tau_{n-1}} \circ \cdots  \circ J_{\pi_{1}} \, \circ U_{\tau_0} \, (\brho_0) }{\Tr[\dots]}
\\ 
\Theta_n 
&:=
\Theta(t_n, \omega)
\end{aligned}
\end{equation}

We call the set of states appearing in a unravelling $\omega \in \Unravel$ 

\begin{equation}
\begin{aligned}
\Omega_{\brho_0}(\omega)
:&=
\{\brho_0 = \Theta_0\} \cup
\left\{
\frac{J_{\pi_{n}} \, \circ U_{\tau_{n-1}} \circ \cdots  \circ J_{\pi_{1}} \, \circ U_{\tau_0} \, (\brho_0) }{\Tr[\dots]} \, : \, n \in \N, \, \omega= (t_i, \pi_i)_{i \in \N}
\right\} = \\
&= 
\left\{
\Theta_n(\omega)  \, : \, n \in \N_0, \, \omega= (t_i, \pi_i)_{i \in \N}
\right\} \, ,
\end{aligned}
\end{equation} 
and the sequence $\left(\Theta_n\right)_{n \in \N} = \left(\Theta_{\brho_0}(t_n, \omega)\right)_{n \in \N}$ the \textit{discrete quantum trajectory}.

\begin{figure}[H]
\begin{center}
\includegraphics[width=0.7\columnwidth]{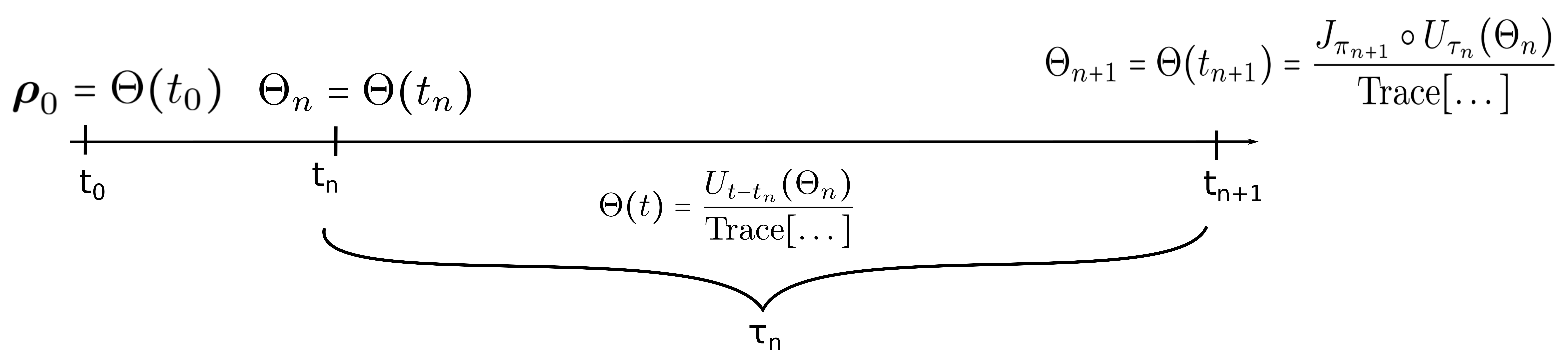} 
\caption{ Illustration of a quantum trajectory    }
\label{RelaxingLindblad_Illustration_unravelling} 
\end{center}
\end{figure}

Such a process is called \textit{renewal process \cite{grimmett2020probability,douc2018markov,bremaud2013markov}}.

\subsection{Ket-state unravellings} \label{KetStateUnravellings}

If some state  in the quantum trajectory is a pure state $( \Theta_n =\, P_{\psi_n}:= \ket{\psi}\bra{\psi} \hat{=} \, \psi_n )$, the unravelling reduces to a ket-state unravelling from then on, and equations \eqref{WaitingTimeDistribution}, \eqref{Prob_WaitingTime}, \eqref{ProbForNextJump}, and \eqref{StatesInAQuantumTrajectory} become

\begin{equation} \tag{\ref{WaitingTimeDistribution}'}
\begin{aligned}
f(t \, | \, \psi) 
=
- \Tr[U_t (P_\psi )]
=
- \Tr[
P_{U_t \, ( \psi) }
]
=
- \| 
U_t \, \psi
\|^2
\end{aligned}
\end{equation}

\begin{equation} \tag{\ref{Prob_WaitingTime}'}
\begin{aligned}
\Prob \, 
\bigl( 
\tau_n \in [0, T) \, | \, \psi_n
\bigr)
=
\int_0^T  \d t
\underbrace{
\hspace*{5mm} 
f (t \,| \, \psi)
\hspace*{5mm}}_{ - \frac{d}{dt} \|U_T ({\psi_n}) \|^2 } = 
1 - \|U_T ({\psi_n}) \|^2
\end{aligned}
\end{equation}

\begin{equation} \tag{\ref{ProbForNextJump}'}
\begin{aligned}
\Prob \, \bigl(\pi_n = k  \,|\, \psi_n,  \, \tau=\tau_n\bigr) 
=
\frac{f^{(k)}(\tau_n \,|\, \psi_n)}{\sum\limits_{j\in J } f^{(j)}(\tau_n\,|\, \psi_n)} 
\end{aligned}
\end{equation}

The states occurring during a ket state unravelling $ \omega \in \Unravel$ are given by

\begin{equation} \tag{\ref{StatesInAQuantumTrajectory}'}
\begin{aligned}
\text{  for } {\color{black}t} \in [t_{n}, t_{n+1}):
\Psi_{\psi_0}({\color{black}t}, \omega) 
&= 
\frac{U_{{\color{black}t}-t_{n}} \, (\Psi_n)}{\|\dots\|}
= 
\frac{U_{{\color{black}t}-t_{n}} \circ J_{\pi_{n+1}} \, \circ U_{\tau_n} \circ \cdots  \circ J_{\pi_{1}} \, \circ U_{\tau_0} \, (\Psi_0) }{\|\dots\|}
\\ 
\Psi_n 
&:=
\Psi(t_n, \omega)
\end{aligned}
\end{equation}

\end{itemize}

The transition to a ket-state unravelling occurs for instance when a Lindblad operator of rank one $(V=\ket{\psi}\bra{\varphi} )$  is applied at some point in the quantum trajectory,

\begin{equation}
\begin{aligned}
\Theta_{n+1} 
=
\frac{ \ket{\psi}\bra{\varphi} \, 
U_{\tau_n}( \Theta_n) \, \ket{\varphi}\bra{\psi} }{ \Tr\left[\dots  \right] } = \ket{\psi}\bra{\psi} 
\hat{=}
\psi. 
\end{aligned}
\end{equation}

So, in some sense, Lindblad operators of rank one are "quasi-classical". 

\section{ Computing the stationary solution of the Lindblad equation based on quantum-jump unravelling}\label{ComputingTheStationarySolutionOfTheLindbladEquationBasedOnQuantum-jumpUnravelling}

The stationary solution $\brho^\infty(\brho_0)$ for an initial state $\brho_0$ is obtained by taking the long-time average of expression \eqref{rho_unravel}
and interchanging the time average with the ensemble average,
\begin{equation}
\brho^\infty(\brho_0)
=
\int\limits_{\Unravel} \,
\Theta_{\brho_0}^\infty(\omega)\, \d \Prob_{\brho_0}(\omega), 
\end{equation}
where the time average for a single trajectory is given by 
\begin{equation}
\Theta_{\brho_0}^\infty(\omega)
=
\langle \Theta_{\brho_0}(t, \omega) \rangle_{t \geq 0}
=
\lim\limits_{T \to \infty}
\frac{1}{T} \int\limits_0^T \Theta_{\brho_0}(t, \omega) \, \d \, t
.     
\end{equation}

We therefore need to calculate the time averages of quantum trajectories and their probabilites. 
Instead of computing the probability density function  $\Prob(\omega)$ for all $\omega \in \Unravel$ separately, we will divide the set of all possible unravellings into a finite partition  $\Unravel = \dot{\bigcup\limits}_{i=1}^n \, \Unravel(B_i)$ of $n \in \N$ disjoint measurable sets, such that $\Unravel(B_i)$ contains all quantum trajectories with the same long-term behaviour 
$\Theta_{B_i}$,
\begin{equation}
\begin{aligned}\label{EnsembleAverage_over_MinimalAbsorbingSet}
\brho^\infty(\brho_0)
=
\sum\limits_{i=1}^n \;\, \int\limits_{\Unravel(B_i)} 
\underbrace{
\Theta_{\brho_0}^\infty(\omega)
}_{
\Theta_{B_i}
}\, \d \Prob_{\brho_0}(\omega)
=
\sum\limits_{i=1}^n
\Prob\Bigl(\Unravel(B_i) \,\bigl|\, \brho_0\Bigr) \, \Theta_{B_i}. 
\end{aligned}
\end{equation}
Calculating the different building blocks of this expression is the goal of the next subsections.

An explicit expression for the time average  $\Theta_{\brho_0}^\infty(\cdot)$ of a single quantum trajectory will be obtained in section \ref{TimeAverageOfASingleQuantumTtrajectory}. To find the different possible long-time behaviors of quantum trajectories, we view an unravelling as a discrete-time Markov chain in section \ref{DiscreteTimeMarkovChainsWithFiniteStateSpace}, where the next state $\Theta_{n+1}$ of the discrete quantum trajectory depends only on the previous state $\Theta_n$. 
In section \ref{ConstructingTheMarkovChainForQuantumTrajectories} we explicitly construct the set of possible states in the Markov chain 
and the matrix of transition probabilities, and we illustrate this procedure by three examples in section \ref{ExamplesOfDiscreteTimeMarkovChainsInTheContextOfQuantumJumpUnraveling}. 
This gives us both the stationary states of the Markov chain and the probabilities 
for the different long-term behaviours. 

In section \ref{Evaluating_Theta_BForAGeneralConditionalHamiltonian_H_c}, we will evaluate the time averaged states between jumps. 
By combining all these results, we can obtain in section \ref{TheStationarySolutionOfTheLindbladEquation} the final expression for the stationary solutions of the Lindblad equation \eqref{Lindbladian}.

\subsection{ Time average of a single quantum trajectory }\label{TimeAverageOfASingleQuantumTtrajectory}

The time average of a single quantum trajectory is given by
\begin{equation}\label{IntegalExpressionOfTheta}
\begin{aligned}
\Theta_{\brho_0}^\infty (\omega)
=
\lim\limits_{T \to \infty}
\frac{1}{T}
\int_0^T 
\Theta_{\brho_0}(t, \omega)  \, \d \,t
=:
\langle \Theta_{\brho_0}(t, \omega) \rangle_{t \geq 0}. 
\end{aligned}
\end{equation}

 according to the theorem quoted in appendix \ref{ChangingLimits_TimeAndEnsembleAverage}.

In order to evaluate this expression further, we denote by 

\begin{equation}
\begin{aligned}
J(T) 
&:=
\left|\{ n \in \N \,:\, t_n \leq T \}\right|  \text{  the number of jumps before the time $T>0$  ,  and and by } \\
J_s(T) 
&:=
\left|\{ n \in \{1, \dots, J(T)\} \, :\, \Theta_n = \Theta_s \} \right| \text{  the number of times the state $\Theta_s$ } \\ 
&  \text{                  has appeared in the quantum trajectory before the time $T$.}
\end{aligned}
\end{equation}

We break the time  interval $[0,T)$ into a disjoint union of smaller time
intervals $[t_{n}, t_{n+1})$ between jumps, 
\[
[0,T) = \dot{\bigcup\limits}_{n=0}^{J(T)-1} \; [t_{n}, t_{n+1}) \dot{\cup} \; [t_{J(T)}, T)\, ,
\]
and compute

\begin{equation}\label{Theta_infty_General}
\begin{aligned}
\Theta_{\brho_0}^\infty(\omega) 
&\xlongequal{\text{theorem }\eqref{ChangingLimits_TimeAndEnsembleAverage}}
\lim\limits_{T \to \infty} \frac{1}{T} \int_0^T  \;  \Theta_{\brho_0} (t, \omega) \,d  \, t = \\
&= \lim\limits_{T \to \infty} \frac{1}{T} \sum\limits_{n=0}^{J(T)-1 } 
\int_{t_{n}}^{t_{n+1}} \, 
\frac{ U_{t-t_{n}}(\, \Theta_n \,)
}{ \Tr\left[ \dots \right] } 
\, d \, t
+ 
\lim\limits_{T \to \infty} \frac{1}{T} I_{\Theta_{J(T)}} (T-t_{J(T)})
= \\ 
&= \lim\limits_{T \to \infty} \frac{1}{T} \sum\limits_{n=0}^{J(T)-1 } 
\underbrace{
\int_0^{\tau_n} \, 
\frac{ U_{t}(\, \Theta_n \,)
}{ \Tr\left[ \dots \right] } 
\, d \, t
}_{
I_{\Theta_n} \left( \tau_n \right)
}
+ 
\lim\limits_{T \to \infty} \frac{I_{\Theta_{J(T)}} (T-t_{J(T)})}{T}  \\
&=
\lim\limits_{T \to \infty} 
\underbrace{
\left( 
\frac{1}{T} 
\sum\limits_{n=0}^{J(T)-1 } 
I_{\Theta_n} ( \tau_n )
\right)
}_{
\text{'jump part'}
}
+ 
\lim\limits_{T \to \infty}
\underbrace{
\left( 
\frac{1}{T} \, I_{\Theta_{J(T)}} (T-t_{J(T)})
\right), 
}_{
\text{'remainder part'}
}
\end{aligned}
\end{equation}

where the time-weighted average $I_{\Theta}(\tau)$ was defined in equation \eqref{Def_I_ThetaOfTau}.

The number of quantum jumps can be finite or infinite, so in the following we have to distinguish these two cases. 
As we will see, the 'jump part' vanishes in the case of finitely many jumps and the time average will be determined by the 'remainder part', and vice versa when the number of jumps is infinite.

\subsubsection{The number of quantum jumps is finite}

When the number of quantum jumps is finite, we can set $J(T) \xlongequal{T\text{  large enough}} J_{\text{Max}} =: J_M$. So the discrete quantum trajectory will stop at the state $\Theta_\text{trap} := \Theta_{J_M}$. 
As mentioned before, this can only happen if $ \int_{\R_{\geq \, 0}} f(t\,|\, \Theta_\text{trap}) \, \d t < 1$.

We claim that the time average of this particular quantum trajectory equals $\Theta_{\brho_0}^\infty = \lim\limits_{T \to \infty} \, 
\frac{1}{T} \, 
\int_0^T U_t(\Theta_\text{trap}) \, \d t$. 

\begin{proof}

\begin{itemize}


\item[a)]
The 'jump part': \\

\begin{equation}
\begin{aligned}
\left\|
\frac{1}{T}
\sum\limits_{n=0}^{J(T)-1}
\underbrace{
I_{\Theta_n}(\tau_n)
}_{
\leq \, \tau_n \, \sup\limits_{t \in [0, \tau_n]} \, \|I_{\Theta_n}(t)\|
}
\right\|
\leq \,
\max\limits_{n \in \{0, \dots, J_M-1\}} \sup\limits_{t \in [0, \tau_n]} \|I_{\Theta_n}(t)\|
\, 
\frac{t_{J_M}}{T}
\xlongrightarrow{T \to \infty}
0. 
\end{aligned}
\end{equation}


\item[b)]
The 'remainder part': \\

\begin{equation}
\begin{aligned}
\frac{I_{\Theta_{J(T)}}(T - t_{J(T)})}{T}
&=
\frac{I_{\Theta_{J_M}}(T - t_{J_M})}{T - t_{J_M}} \, 
\overbrace{
\left(\frac{T - t_{J_M} }{T}\right)
}^{
\xlongrightarrow{T \to \infty}1
} 
%
\xlongrightarrow{T \to \infty}
\lim\limits_{T \to \infty}
\frac{1}{T}
I_{\Theta_\text{trap}}(T) = \\
&=
\lim\limits_{T \to \infty} \, 
\frac{1}{T} \, 
\int_0^T \frac{U_t(\Theta_\text{trap})}{Tr[]} \, \d t
=\langle \Theta_\text{trap}(t)\rangle_{t \geq \, t_{J(T)}}. 
\end{aligned}
\end{equation}

\end{itemize}

\end{proof}

\subsubsection{The number of quantum jumps is infinite}
When the number of quantum jumps is infinite, the discrete quantum trajectory does not stop. 
We show that the 'remainder part' vanishes and derive an expression for the 'jump part'. 

\begin{itemize}
\item[a)] The 'remainder part': 

The 'remainder' term $\lim\limits_{T \to \infty} \frac{I_{\Theta_{J(T)}} (T-t_{J(T)})}{T}$ tends to zero when $T$ approaches infinity, 

\begin{equation}
\begin{aligned}
\left\|
\frac{I_{\Theta_{J(T)}} (T-t_{J(T)})}{T} 
\right\|
&=
\left\|
\frac{I_{\Theta_{J(T)}} (T-t_{J(T)})}{T-t_{J(T)}}
\, 
\left(
\frac{T-t_{J(T)}}{T}
\right)
\right\|
\, \leq  \\ 
& \leq 
\sup\limits_{n \in \N}
\sup\limits_{t \in [0, \tau_n]}
\left\|
\frac{ U_t( \, \Theta_n \,) }{ \Tr\left[ \dots  \right] }
\right\|
\;
\, 
\left(
1-\frac{t_{J(T)}}{T}
\right)
\xlongrightarrow{T \to \infty}
0, 
\end{aligned}
\end{equation}

where we used the fact that $\lim\limits_{T \to \infty}\frac{t_{J(T)}}{T}$ = 1 (see appendix \ref{t_J(T)_DividedBy_T_TendsTo_1}).

\item[b)] The 'jump part': 

For a fixed quantum trajectory $\omega \in \Unravel$ that does not stop we denote by $\alpha_s(k) \in \N$ the position in that quantum trajectory where the corresponding state $\Theta_s \in \Omega_{\brho_0}(\omega)$ appears for the $k$-th time, so

\begin{equation}\label{DefinitionsForalpha_s}
\begin{aligned}
\alpha_s(k) 
&:=
\min\limits_{M\in\N} \Bigl\{\, |\{ \Theta_n \,:\, n\leq M, \Theta_n = \Theta_s \}|=k  \, \Bigr\} . 
\end{aligned}
\end{equation}

Note that we define $\alpha_s(k)$ only for those states that actually appear in $\omega \in \Unravel$.

In particular, we have $\Theta_{\alpha_s(k)} = \Theta_s$ for all $s\in \s$ and for all $k\in \N$, and the waiting time $\tau_{\alpha_s}(k)$ has the probability density function $f(\cdot \,|\, \Theta_{\alpha_s(k)}) = f(\cdot \,|\, \Theta_s)$ (see equation \eqref{WaitingTimeDistribution} for a definition).

The sums of the form $\frac{1}{J(T)}\sum\limits_{n=0}^{J(T)-1} x_n$, with $x_n \in \{\tau_n, I_{\Theta_n}(\tau_n)\}$ in the 'jump part' of equation \eqref{Theta_infty_General} can be rearranged to 
\begin{equation}
\begin{aligned}
\frac{1}{J(T)}\sum\limits_{n=0}^{J(T)-1} x_n
=
\sum\limits_{s \in \s} \, 
\frac{J_s(T)}{J(T)} \, 
\frac{1}{J_s(T)} \, 
\sum\limits_{k=1}^{J_s(T)} x_{\alpha_s(k)}. 
\end{aligned}
\end{equation}

For all states $s \in \s$, both the waiting times $\{\tau_{\alpha_s(k)} \,:\, k\in \N\}$ and the time-weighted  average states   
$\{I_{\Theta_s} \left(\tau_{\alpha_s(k)}\right) \,:\, k\in \N\}$
are independent and identically distributed, and by the law of large numbers we have 
%
\begin{equation}\label{LawOfLargeNumbers_Applied}
\begin{aligned}
\frac{1}{K}\, \sum\limits_{k=0}^{K-1} \tau_{\alpha_s(k)} &\xlongrightarrow{K\to \infty} 
\overline{\tau_s}, \\
\frac{1}{K} \sum\limits_{k=0}^{K-1} \, I_{\Theta_s} \left( \tau_{\alpha_s(k)}  \right)
  &\xlongrightarrow{K\to\infty}
\,  \int_0^\infty I_{\Theta_s} (\tau) \, f(\tau \,|\, \Theta_s) \, d \, \tau
 =: 
 \overline{I_s(\tau_s)}, \\
\text{   pointwise, almost surely. }
\end{aligned}
\end{equation}
The term $\frac{J_s(T)}{J(T)}$ can be interpreted as the fraction of jumps in the quantum trajectory whose outcome yields the state $\Theta_s$. 

For long times, the trajectory will eventually be in a minimal absorbing set $B \subseteq \Omega_{\brho_0}(\omega)$, which is a subset of states that is both \textit{absorbing} (which means that there are no transitions out of this set) and \textit{minimal} in the sense of set inclusions (when $B$ is a minimal absorbing set and $A \subseteq B$ is absorbing, we have $A=B$), (see section \ref{DiscreteTimeMarkovChainsWithFiniteStateSpace} and appendix \ref{TrajectoriesAreCapturedByMinimalAbsorbingSets}. Therefore, the fraction $\frac{J_s(T)}{J(T)}$  of jumps landing in state $\Theta_s$ converges to a stationary value $q(\Theta_s \,|\, B)$ associated with this minimal absorbing set, 

\begin{equation}\label{Probability_P_s}
\begin{aligned}
\frac{J_s(T)}{J(T)}
&=
\frac{1}{J(T)}
\sum\limits_{k=0}^{J(T)-1}
\delta_{\Theta_n, \Theta_s}
\xlongrightarrow{T \to \infty}
q(\Theta_s \,|\, B), 
\end{aligned}
\end{equation}
with $q(\Theta_s \,| \, B)$ being the \textit{relative frequency} with which the state $\Theta_s\in \Omega_{\brho_0}(\omega)$ appears in the discrete quantum trajectory $(\Theta_n)_{n \in \N_0}$. Since the numbers $\left(q(\Theta_s \,| \, B)\right)_{s \in \s}$ are non-negative and add up to one, we can interpret $q(\Theta_s \,| \, B)$ as the  probability of state $\Theta_s$, provided the discrete quantum trajectory is eventually in the minimal absorbing set $B \subseteq \Omega_{\brho_0}(\omega)$. 

This idea will be picked up in section \ref{DiscreteTimeMarkovChainsWithFiniteStateSpace}.

If the probability for a state $\Theta_s \in \Omega_{\brho_0}(\omega)$ in a quantum trajectory is strictly positive, then it must appear infinitely often in the corresponding trajectory: 
$q(\Theta_s \,|\, B) =  \lim\limits_{T \to \infty}  \frac{J_s(T)}{J(T)} >0 \Longrightarrow  \lim\limits_{T \to \infty} \, J_s(T) = \infty$. 

Now we can compute the 'jump part'


\begin{equation}
\begin{aligned}
\lim\limits_{T \to \infty}\frac{1}{T}\sum\limits_{n=0}^{J(T)-1 } 
 I_{\Theta_n} \left( \tau_n \right)
%
&=
\lim\limits_{T \to \infty} \frac{\frac{1}{J(T)} \, \sum\limits_{n=0}^{J(T)-1 } 
 I_{\Theta_n} \left( \tau_n \right)}{\frac{1}{J(T)} \, \sum\limits_{n=0}^{J(T)-1 } 
  \tau_n + \left(\frac{T - t_{J(T)}}{J(T)}\right)} = \\
%
&= 
\lim\limits_{T \to \infty} \frac{\sum\limits_{s \in \s}\frac{J_s(T)}{J(T)}\, \frac{1}{J_s(T)} \sum\limits_{k=0}^{J_s(T)-1 }  I_{\Theta_{\alpha_s(k)}} \left(\tau_{\alpha_s(k)} \right)}{\sum\limits_{s \in \s} \frac{J_s(T)}{J(T)}\, \frac{1}{J_s(T)}\sum\limits_{k=0}^{J_s(T)-1 }  \tau_{\alpha_s(k)} + \left(\frac{T - t_{J(T)}}{J(T)}\right)} 
 = \\ \\
&\xlongequal{|\s|<\infty} 
\frac{
\sum\limits_{s \in \s} 
\overbrace{
\left( \lim\limits_{T \to \infty} \frac{J_s(T)}{J(T)}\right)
}^{
q(\Theta_s \,|\, B) 
}\, 
\overbrace{
\left( \lim\limits_{T \to \infty} \frac{1}{J_s(T)} \sum\limits_{k=0}^{J_s(T)-1 }  I_{\Theta_{\alpha_s(k)}} \left(\tau_{\alpha_s(k)} \right)\right)
}^{
\overline{I_s(\tau_s)}
}
}
{\sum\limits_{s \in \s} 
\underbrace{
\left( \lim\limits_{T \to \infty} \frac{J_s(T)}{J(T)}\right)
}_{
q(\Theta_s\,|\, B) 
}\, 
\underbrace{
\left( \lim\limits_{T \to \infty} \frac{1}{J_s(T)}\sum\limits_{k=0}^{J_s(T)-1 }  \tau_{\alpha_s(k)}\right)
}_{
\overline{\tau}_s
} + 
\underbrace{
\left( \lim\limits_{T \to \infty} \left(\frac{T - t_{J(T)}}{J(T)}\right)\right)
}_{0}} \\
&\xlongequal{\eqref{LawOfLargeNumbers_Applied}\,+\,\eqref{Probability_P_s}}
\frac{\sum\limits_{s \in \s}q(\Theta_s \,|\, B)  \, 
\overline{I_s(\tau_s)}
}{\sum\limits_{s \in \s}q(\Theta_s\,|\, B)  \, \overline{\tau}_s}
=
\frac{\sum\limits_{s \in \s}q(\Theta_s\,|\, B)  \, 
\overline{I_s(\tau_s)}
}{\Tr[\cdots]}
=: \Theta_{B}. 
\end{aligned} 
\end{equation}

\end{itemize}

\subsubsection{Summarizing the two cases}\label{TheGeneralCase_WhenTheNumberOfQuantumJumpsIsArbitrary}

When the number of quantum jumps is infinite, this minimal absorbing set is a subset of the possible states of the discrete quantum trajectory, $B\subseteq \bigcup_{\omega \in \Unravel}\Omega_{\brho_0}(\omega)$.
When on the other hand the number of quantum jumps is finite, that is when the discrete quantum trajectory ends at a possible trapping state $\Theta_\text{trap}$, then 
the discrete quantum trajectory also ends in a minimal absorbing set, which is in this case
given by $B = \bigl\{\lim\limits_{T\to \infty}\, \frac{1}{T} I_{\Theta_\text{trap}}(T)\bigr\}$.

Taking the two expressions for the average of the quantum trajectory together, we obtain

\begin{equation}\label{Theta_B}
\begin{aligned}
\Theta_{B}
:=
\Theta_{\brho_0}^\infty(\omega)
=
\begin{cases}
\lim\limits_{T \to \infty}
\frac{1}{T} \, 
I_{\Theta_{J_M}}(T)
, &\text{  for finitely many jumps} \\ \\
\frac{\sum\limits_{s \in \s}q(\Theta_s \,|\, B )  \, 
\overline{I_s(\tau_s)}
}{\Tr[\cdots]}
,&\text{ when the number of jumps is infinite. }
\end{cases}
\end{aligned}
\end{equation}

This result shows that the long-term behaviour of a quantum trajectory does not depend on the trajectory $\omega \in \Unravel$ itself, but only on the minimal absorbing set $B$ it is eventually captured in.

\subsection{Discrete-time Markov chains with finite state space}\label{DiscreteTimeMarkovChainsWithFiniteStateSpace}

Now that we have computed an expression the time-averaged state $\Theta_{\brho_0}^\infty(\omega)$ for a fixed $\omega \in \Unravel$, we need to evaluate two quantities in order to obtain the stationary states of the Lindblad equation: 

\begin{itemize}
\item[(i)] First, the probability $\Prob(\Unravel(B) \, | \, \brho_0)$ for a quantum trajectory to land in a minimal absorbing set $B$. 
\item[(ii)] Second, the expression for the stationary states $\Theta_B$, corresponding to the minimal absorbing state  $B$. 
\end{itemize}

Depending on the nature of the minimal absorbing set, we need for the expression of the stationary state either

\begin{itemize}
\item[(iia)] the probabilities $q(\Theta_s \,|\, B)$ for the states $\Theta_s \in B$ \textit{within} the minimal absorbing set $B$ as well as their time-weighted ensemble average $\overline{I_s(\tau_s)}$ or 

\item[(iib)] the long-term average $\lim\limits_{T \to \infty}
\frac{1}{T} \, 
I_{\Theta_\text{trap}}(T)
$, where $\Theta_\text{trap}$ is a possible trapping state. 
\end{itemize}

\begin{figure}[H]
\centering
\includegraphics[width=0.4\columnwidth]{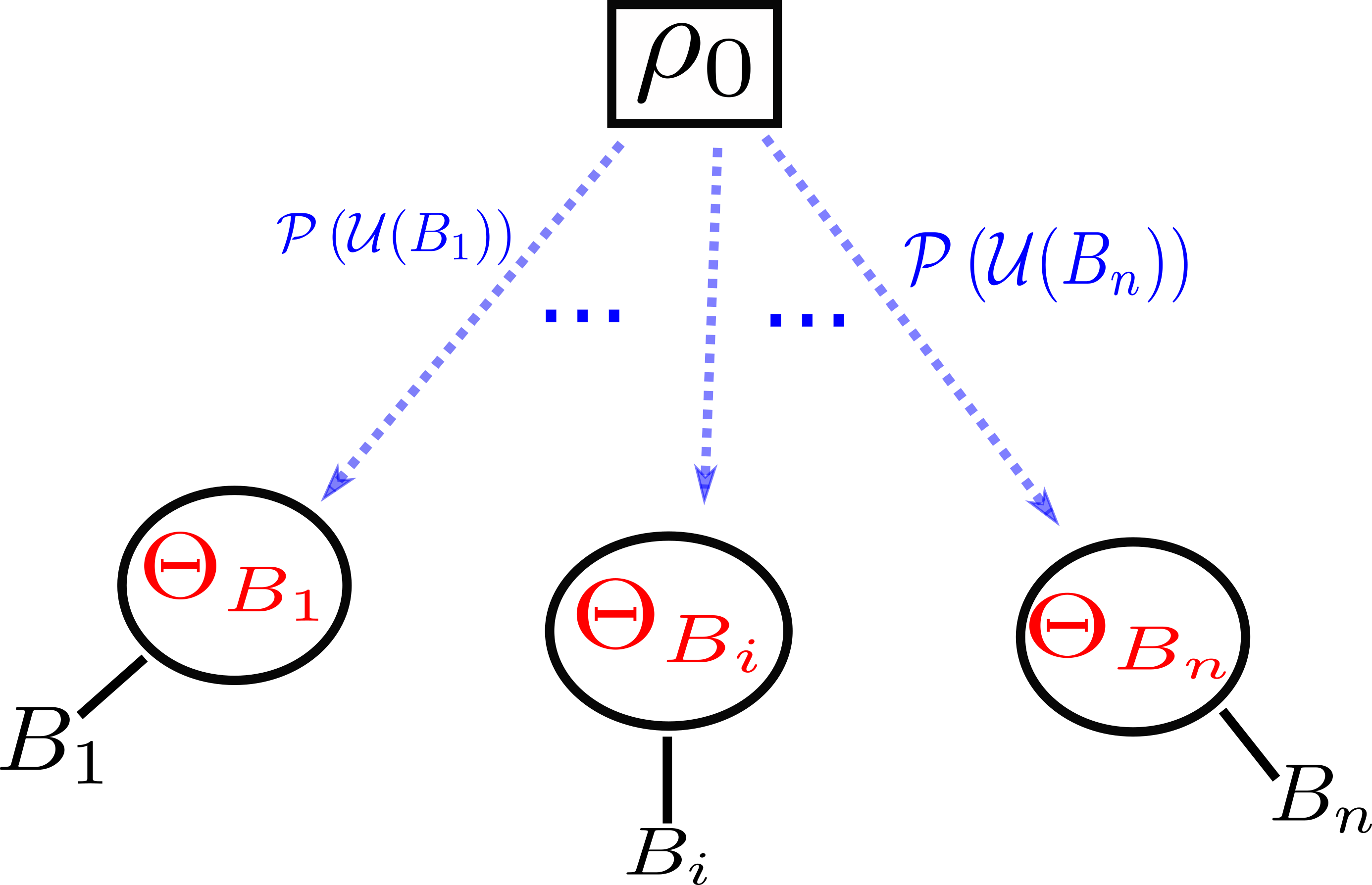}
\caption{Illustrating the significance of a Markov chain in a quantum state unraveling: In order to compute the stationary state we need both the long-term behaviour \begin{color}{red}$\Theta_{B_i}$\end{color} of a single quantum trajectory contained in the minimal absorbing set $B_i$, as well as the probability \begin{color}{blue}$\Prob(\Unravel(B_i))$\end{color} for such a trajectory to be captured by this minimal absorbing set $B_i$, where $i \in \{1, \dots, n\}$. }
\label{Illustrating_NeedForMarkovChain}
\end{figure}

Both the probability for reaching each minimal absorbing set $B \subseteq \Omega_{\brho_0}$,  $\Prob(\Unravel(B)) \,|\, \brho_0)$,  as well as the probability for each state  $q(\Theta_s \,|\, B)$ within such a minimal absorbing set
 can be obtained by considering only the states $\Theta_s \in \Omega_{\brho_0}$ of the discrete quantum trajectory and the transition probabilities between them. The discrete quantum trajectory is a Markov chain, since the probability for the next state $\Theta_{n+1}$ depends only on the previous state $\Theta_n$, 
 
\begin{equation}
\begin{aligned}
\Prob(\Theta_{n+1} \,|\, \Theta_n, \Theta_{n-1}, \dots, \Theta_0)
=
\Prob(\Theta_{n+1} \,|\, \Theta_n). 
\end{aligned}
\end{equation}

We will therefore in the following summarize those results for classical discrete-time Markov chains that will be relevant for obtaining  the two types of probabilities. The connection to the quantum trajectories and the evaluation of the time averages between jumps will follow in the subsequent subsections.

We consider a classical, discrete-time Markov process $(X_n)_{n \in \N_0}$ taking values in the finite state space $\Omega$. The transition probabilities are defined via 

\begin{equation}
\begin{aligned}
Q_{ij}
:=
q_{j \to i}
:=
\Prob\left(
X_{n+1} = i \,|\, X_n = j
\right)
=
\Prob\left(
X_{n+1} = i \,|\, X_n = j, \dots, X_0 = i_0
\right). 
\end{aligned}
\end{equation}

This specifies the transition matrix $Q \in \left(\R_{\geq \, 0} \right)^{|\Omega| \times |\Omega|}$, which is a column-stochastic matrix that satisfies

\begin{equation}
\begin{aligned}
Q_{ij} &\geq 0 \text{     , for all } i,j \in \Omega \text{  and } \\
\sum\limits_{i = 1}^{|\Omega|} Q_{ij} &= 1 \text{     , for all } j \in \Omega  . 
\end{aligned}
\end{equation}

The states and the nonzero transition probabilities form a network $(\Omega, \Edge)$, with the states $\Omega$ being the nodes and the nonzero transitions $\Edge \subseteq \Omega \times \Omega$ being the directed links between them. 
When there is no transition from state $i$ to state $j$, the associated transition probability $q_{i \to j}$ vanishes. 
The transition probability indicate the strength of the link.

For each finite-size Markov chain there is an associated state transition network (a directed weighted graph). An example is given in Figure \ref{Example_StateTransitionNetwork_MarkovChain}: 

\begin{figure}[H]
\centering
\includegraphics[width=0.6\columnwidth]{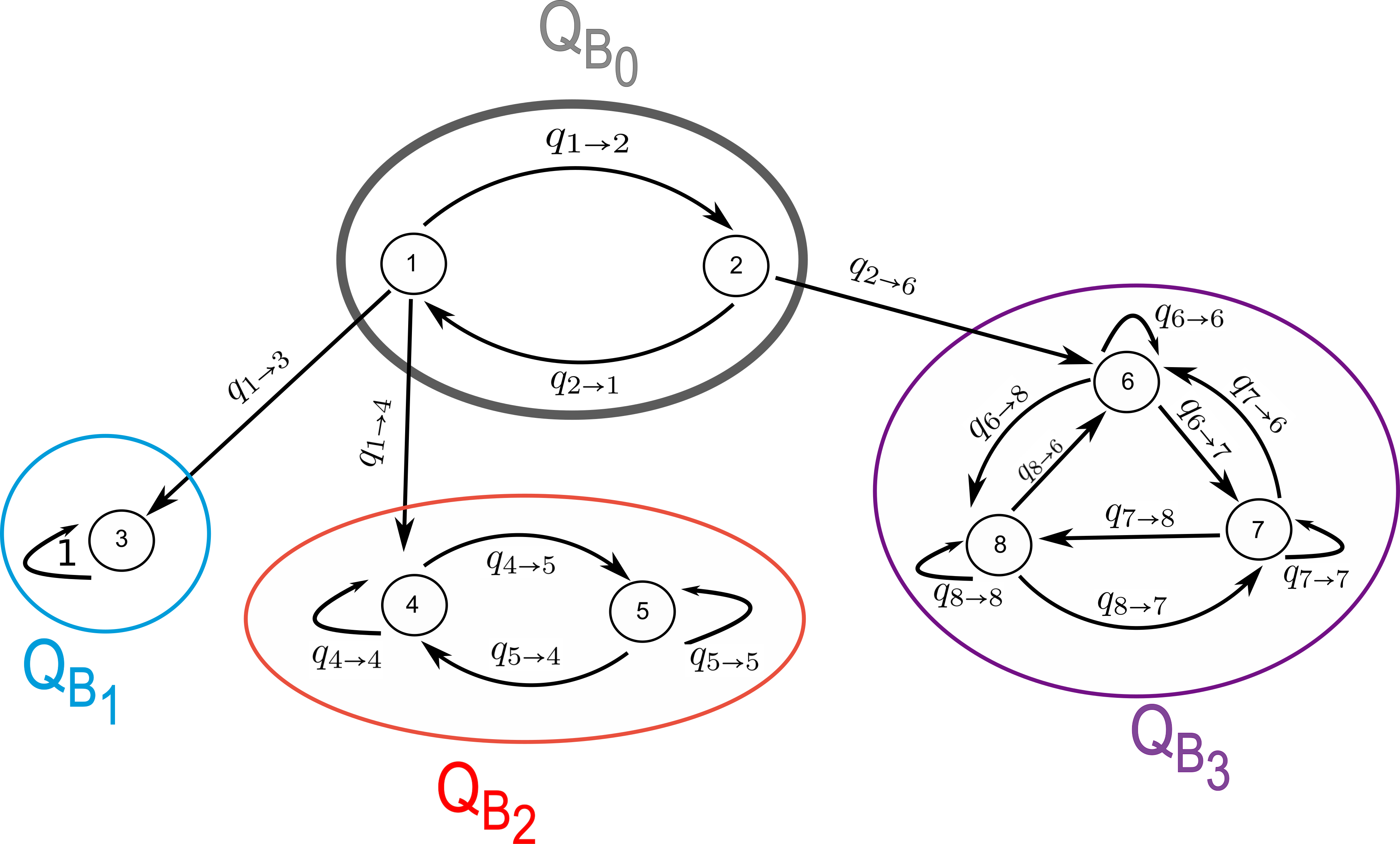}
\caption{Example of a state transition network of the Markov chain with the minimal absorbing sets
${\begin{color}{TUDa_1b}   B_1 \end{color}}$, 
${\begin{color}{TUDa_9b}   B_2 \end{color}}$,  
${\begin{color}{TUDa_11b}  B_3 \end{color}}$ and the transition matrix \\ \\
\hspace*{25mm}
$Q = \begin{pmatrix}
0 & q_{2 \to 1} & 0 & 0 & 0 & 0 & 0 & 0 \\
q_{1 \to 2} & 0  & 0  & 0  & 0  & 0  & 0  & 0 \\
q_{1 \to 3} & 0 & {\begin{color}{TUDa_1b}1\end{color}} & 0  & 0  & 0  & 0  & 0  \\
0 & 0 & 0 & {\begin{color}{TUDa_9b}q_{4 \to 4}\end{color}} & {\begin{color}{TUDa_9b}q_{5 \to 4}\end{color}} & 0  & 0  & 0  \\
0 & 0 & 0 & {\begin{color}{TUDa_9b}q_{4 \to 5}\end{color}} & {\begin{color}{TUDa_9b}q_{5 \to 5}\end{color}} & 0  & 0  & 0  \\
0 & q_{2 \to 6} & 0  & 0  & 0  & {\begin{color}{TUDa_11b}q_{6 \to 6}\end{color}}   &  {\begin{color}{TUDa_11b}q_{7 \to 6}\end{color}}   &  {\begin{color}{TUDa_11b}q_{8 \to 6}\end{color}}  \\
0 & 0 & 0  & 0  & 0  & {\begin{color}{TUDa_11b}q_{6 \to 7}\end{color}}   &  {\begin{color}{TUDa_11b} {\begin{color}{TUDa_11b}q_{7 \to 7}\end{color}}   \end{color}}   &  {\begin{color}{TUDa_11b}q_{8 \to 7}\end{color}}  \\
0 & 0 & 0  & 0  & 0  & {\begin{color}{TUDa_11b}q_{6 \to 8}\end{color}}   &  {\begin{color}{TUDa_11b}q_{7 \to 8}\end{color}}  &  {\begin{color}{TUDa_11b}q_{8 \to 8}\end{color}}
\end{pmatrix}$. 
}
\label{Example_StateTransitionNetwork_MarkovChain}
\end{figure}

For all subsets of states $B \subseteq  \Omega$ there is a corresponding subnetwork $(B, \Edge_B)$, where $\Edge_B:= \{(i, j) \in \Edge\, : \, i, j \in  B\}$  consists of all links in the subset $B \subseteq \Omega$. 

We call a subset $B \subseteq \Omega$ \textit{absorbing} if there are no edges pointing out of $B$, that is $q_{i\to j} = 0$ for all $i \in B$ and $j \in B^C$. An absorbing subset $B \subseteq \Omega$ is called \textit{minimal} if for all absorbing subsets $C \subseteq \Omega$ with $C \subseteq B$ we have $C = B$. In
particular, there can be more than one minimal absorbing subset \cite{fernengel2022obtaining}.

We denote with $q_n(i)$ the probability of the system to be in state $i$ at the time step $n \in \N_0$ , with $q_n(i) \geq 0$, $ \sum\limits_{i \in \Omega}q_n(i)=1 $ for all $n \in \N$.

Given an initial probability distribution $\bq_0 \in [0,1]^{|\Omega|}$,  with $\|\bq_0\|_1 = 1 $,   the probability distribution after $n \in \N$ time steps is given by

\begin{equation}
\begin{aligned}
\bq_n (\bq_0)
&=
Q^n \, \bq_0 
=
\sum\limits_{j=1}^{|\Omega|}
q_0(j) \, 
\underbrace{ 
\hspace*{4mm}
Q^n \, \be_j
\hspace*{4mm}
}_{
\left(\Prob(X_n = i \,|\, X_0 = j) \right)_{i \in \Omega}
}
=
\sum\limits_{j=1}^{|\Omega|}
q_0(j)
\colvec{3}
{\Prob(X_n = 1 \,|\, X_0 = j) }
{\vdots}
{\Prob(X_n = |\Omega| \,|\, X_0 = j) }, \\
\text{with }
\Prob(X_n = i \,|\, X_0 = j)
&=
\left(Q^n \right)_{ij} 
=
\sum\limits_{ \alpha_{1} \in \Omega} \dots
\sum\limits_{ \alpha_{n-1} \in \Omega} Q_{i,\alpha_{n-1}} \cdot \dotsc \cdot Q_{\alpha_1, j}. 
\end{aligned}
\end{equation}

The probability that a probability starting in state $j\in \Omega$ will be in state $i\in \Omega$ after $n\in \N$ steps is given by the sum over all possible trajectories $\bigl(j=X_0, X_1, \dots, X_n=i \bigr) \in \Omega^n$, namely  

\begin{equation}
\begin{aligned}
\Prob(\omega=(X_n)_{n \in \N_0} \text{ with } X_0 = i \text{ and }X_n = j)
&=
\Prob(X_n = i \,|\, X_0 = j)
=
\sum\limits_{ \alpha_{1} \in \Omega} \dots
\sum\limits_{ \alpha_{n-1} \in \Omega} Q_{i,\alpha_{n-1}} \cdot \dotsc \cdot Q_{\alpha_1, j}. 
\end{aligned}
\end{equation}

A \textit{limiting distribution} $\lim\limits_{K \to \infty} Q^K \, \bq_0$ does not necessarily exist for a discrete-time Markov chain since oscillations are possible. However, the time average $\lim\limits_{K \to \infty} \frac{1}{K} \sum\limits_{k=0}^{K-1}Q^k \, \bq_0$ exists for all finite state space $\Omega$, and for all initial distributions $\bq_0$ and converges to (one of) the \textit{stationary solutions}, which are eigenvectors of the transition matrix $Q$ to the eigenvalue $\lambda=1$, see appendix \ref{TimeAverageOfADiscreteTimeMarkovChain}.

There is a one-to-one correspondence between minimal absorbing sets of the state-transition network and the stationary solutions of a Markov chain \cite{fernengel2022obtaining}: The number of minimal absorbing sets equals the number of (linearly independent) stationary solutions, and for every minimal absorbing set we can construct an associated eigenvector whose non-zero entries correspond to the states in the minimal absorbing network, see equation \eqref{FormulaFor_q_infty} in appendix  \ref{FormOfStationarySolutions}.

The algorithm for obtaining the stationary solution corresponding to the initial condition $\bq_0$ is the following: 
\begin{itemize}
\item [i)]
Find all minimal absorbing sets $B_1, \dots, B_n$ of the network $\Omega$, with $n \in \N, \, n\leq |\Omega|$. 
\item[ii)]
Compute probability vectors  $\bq^\infty (\bq_0 \in B_i)$ for all $i \in \{1, \dots, n\}$ according to formula \eqref{FormulaFor_p_infty} in the appendix, which will be pairwise orthogonal. 
\item[iii)]
Enlarge the orthonormal system $\left(\bq^\infty (\bq_0 \in B_i)\right)_{i \in \{1, \dots, n\}}$ to an ordered basis 
\begin{align*}
\bigl(\bq^\infty (\bq_0 \in B_1), \dots, \bq^\infty (\bq_0 \in B_n), \bh^{(n+1)}, \dots, \bh^{(|\Omega|)} \bigr), 
\end{align*}
where the first $n$ basis vectors are $\bq^\infty (\bq_0 \in B_1), \dots, \bq^\infty (\bq_0 \in B_n)$. 
\item[iv)]
When the basis representation of the initial condition is given by $\bq_0 = \sum\limits_{i=1}^n \mu_{B_i} \, \bq^\infty(\bq \in B_i) + \sum\limits_{i=n+1}^{|\Omega|} \mu_i \, \bh^{(i)}$, then the corresponding stationary solution is given by

\begin{align}\label{StationarySolutionOfDiscreteTimeMarkovChain}
\bq^\infty(\bq_0)
&=
\sum\limits_{i=1}^n \mu_{B_i} \, \bq^\infty(\bq_0 \in B_i)
=
\sum\limits_{i=1}^n \Prob(B_i \,|\, \bq_0) \, \bq^\infty(\bq_0 \in B_i). 
\end{align}

An example is given in appendix  \ref{IllustratingTheAlgorithmForObtainingTheStationaryProbabilitiesOfADiscrete-timeMarkovChain}. 
\end{itemize}

Note that while the focus in \cite{fernengel2022obtaining} was on continuous-time Markov chains, 
the above steps for discrete-time Markov chains follow directly from there. 
Stationary solutions of a discrete-time Markov chain are identical to the stationary solutions of the corresponding continuous-time Markov chain since eigenvectors to the eigenvalue $\lambda=1$ of transition matrix $Q$ of the discrete-time Markov chain are eigenvalues to the eigenvalue $\lambda=0$ of transition matrix $\G_Q = Q - \Id$ of the associated continuous-time Markov chain: $\G_Q \, \bq^\infty = \bzero = (Q - \Id) \, \bq^\infty$.


\subsection{ Constructing the Markov chain for quantum trajectories}\label{ConstructingTheMarkovChainForQuantumTrajectories}

In order to apply the theory of Markov chains, we need to determine both the states and the transition probabilities. Unlike in classical Markov chains, it can happen that an unravelling stops, as the column sum of the transition matrix need not be one. 
This is the case for possible trapping states. 
When constructing the matrix $Q$, this has to be taken into account. In the following, we will first compute the set of possible states of the Markov chain, before we derive expressions for the transition probabilities.

\subsubsection{The states of the Markov chain}\label{TheStatesOfTheMarkovChain}

We denote the set of possible states in the Markov chain by 
\begin{equation}
\begin{aligned}
\{ \Theta_s \,:\, s \in \s\} 
:=
\Omega_{\brho_0}
&=
\bigcup_{\omega \in \Unravel} \Omega_{\brho_0}(\omega) \, \cup \, \left\{\lim\limits_{T \to \infty} \, \frac{I_{\Theta_\text{trap}}(T)}{T}  \;:\; \Theta_\text{trap} \in \bigcup_{\omega \in \Unravel} \Omega_{\brho_0}(\omega) \text{  and   }
\lim\limits_{t \to \infty} \Tr[U_t(\Theta_\text{trap}) \, \neq \, 0 \right\}. 
\end{aligned}
\end{equation}

\begin{figure}[H]
\begin{center}
\includegraphics[width=0.6\columnwidth]{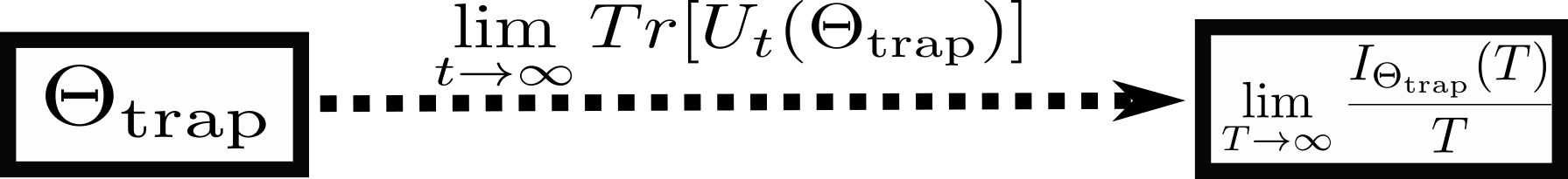} 
\caption{Illustrating a transition from a possible trapping state $\Theta_\text{trap}$ to the associated minimal absorbing set $\left\{\lim\limits_{T \to \infty} \, \frac{I_{\Theta}(T)}{T} \right\}$. The transition probability is given by $\lim\limits_{t \to \infty} \Tr[U_t(\Theta_\text{trap})]>0$.   }
\label{Illustrating_TransitionFromPossibleTrappingState} 
\end{center}
\end{figure}

The last set is only relevant for the possible trapping states $\Theta_\text{trap} $ with $\lim\limits_{t \to \infty} \Tr[U_t(\Theta_\text{trap}) \, \in (0,1]$. In that case, there is a non-zero chance for a transition from the state $\Theta_\text{trap}$ to the minimal absorbing set $\left\{\lim\limits_{T \to \infty} \, \frac{I_{\Theta}(T)}{T} \right\}$ (see figure \eqref{Illustrating_TransitionFromPossibleTrappingState} for an illustration). An explicit expression for the time average of a quantum trajectory that is captured in this minimal absorbing set is given below in section \ref{Theta_B_FinitelyManyQuantumJumps}. 

We choose the initial distribution vector $\bq_{\brho_0}$ to be the first unit vector $\bq_{\brho_0} = \be_1 = (1, 0, \dots, 0) \in \R^{|\s|}$, corresponding to the initial state $\brho_0 = \Theta_0 \in \Omega_{\brho_0}$, which is first element of the set of possible states in the quantum trajectory.

We assume that in all quantum trajectories only finitely many states appear, $|\s|<\infty$.

\subsubsection{The transition probabilities of the Markov chain}\label{TheTransitionProbabilitiesOfTheMarkovChain}

The transition probability 
$\Prob(\Theta_{s_1} \to \Theta_{s_2} )$ is the probability that the next state in the quantum trajectory equals $\Theta_{s_2}$, provided that the last state was $\Theta_{s_1}$. It can be computed to

\begin{equation}\label{Def_TransitionProbability}
\begin{aligned}
\Prob(\Theta_{s_2} \,|\, \Theta_{s_1})
&=
\int_{\R_{\geq \, 0}}
\underbrace{
p(\tau, \Theta_{s_2} \,|\, \Theta_{s_1})
}_{
f(\tau \,|\, \Theta_{s_1}) \, \cdot \, \Prob(\Theta_{s_2} \,|\, \tau, \Theta_{s_1})
}
\d \tau
=
\int_{\R_{\geq \, 0}}
f(\tau \,|\, \Theta_{s_1}) 
\, \cdot \, 
\underbrace{
\Prob(\Theta_{s_2} \,|\, \tau, \Theta_{s_1})
}_{
\sum\limits_{k \in I(s_1 \to s_2)} \, \frac{f^{(k)}(\tau \, | \, \Theta_{s_1})}{f(\tau \, | \, \Theta_{s_1})}
}
\d \tau = \\
&=
\sum\limits_{k \in I(s_1 \to s_2)} \, 
\int_{\R_{\geq \, 0}} 
f^{(k)}(\tau \, | \, \Theta_{s_1})
\d \tau , 
\end{aligned}
\end{equation}

where the sum goes over all Lindblad operators $V_k$ that yield a transition from state $\Theta_{s_1}$ to $\Theta_{s_2}$, or more formally:  $I({s_1} \to {s_2}) :=  \{k \in I \,:\, J_k\circ U_t(\Theta_{s_1})  = \Theta_{s_2} \text{  for some } t>0 \}$.

The following examples illustrate both the states of all possible discrete quantum trajectories as well as the transition probabilities.

\subsection{Examples}\label{ExamplesOfDiscreteTimeMarkovChainsInTheContextOfQuantumJumpUnraveling}

\subsubsection[Example a)]{Example a): $H \propto \Id$,  $V_1 = \begin{pmatrix}
 0 & 0 \\
 1 & 0
 \end{pmatrix}$, $V_2 = \begin{pmatrix}
 0 & 1 \\
 0 & 0
 \end{pmatrix}$. 
}

The set of possible states appearing in the Markov chain can be computed to: 

\begin{equation}
\begin{aligned}
\Omega_{\brho_0}
=
\Bigl\{
\Theta_{s_1} &= \begin{pmatrix}
\brho_{11}(0) & \brho_{12}(0) \\
\brho_{21}(0)& \brho_{22}(0)
\end{pmatrix}, 
\Theta_{s_2} = \begin{pmatrix}
 1 & 0 \\
 0 & 0
\end{pmatrix},
\Theta_{s_3} = 
\begin{pmatrix}
0 & 0 \\
0 & 1
\end{pmatrix}\Bigr\}
\end{aligned}
\end{equation}

The transition matrix equals 

\begin{equation}
\begin{aligned}
Q = \begin{pmatrix}
 0 & 0 & 0 & 0 \\
\brho_{11}(0) & 0 & 1 & 0 \\
\brho_{22}(0) & 1 & 0 & 0 \\
0 & 0 & 0 & 1 
 \end{pmatrix}, 
\text{with the ordered states } 
\left( \Theta_{s_1}=\brho_0, \Theta_{s_2}, \Theta_{s_
3} \right). 
\end{aligned}
\end{equation}

In this example there are no possible trapping states.

\begin{figure}[H]
\begin{center}
\includegraphics[width=0.5\columnwidth]{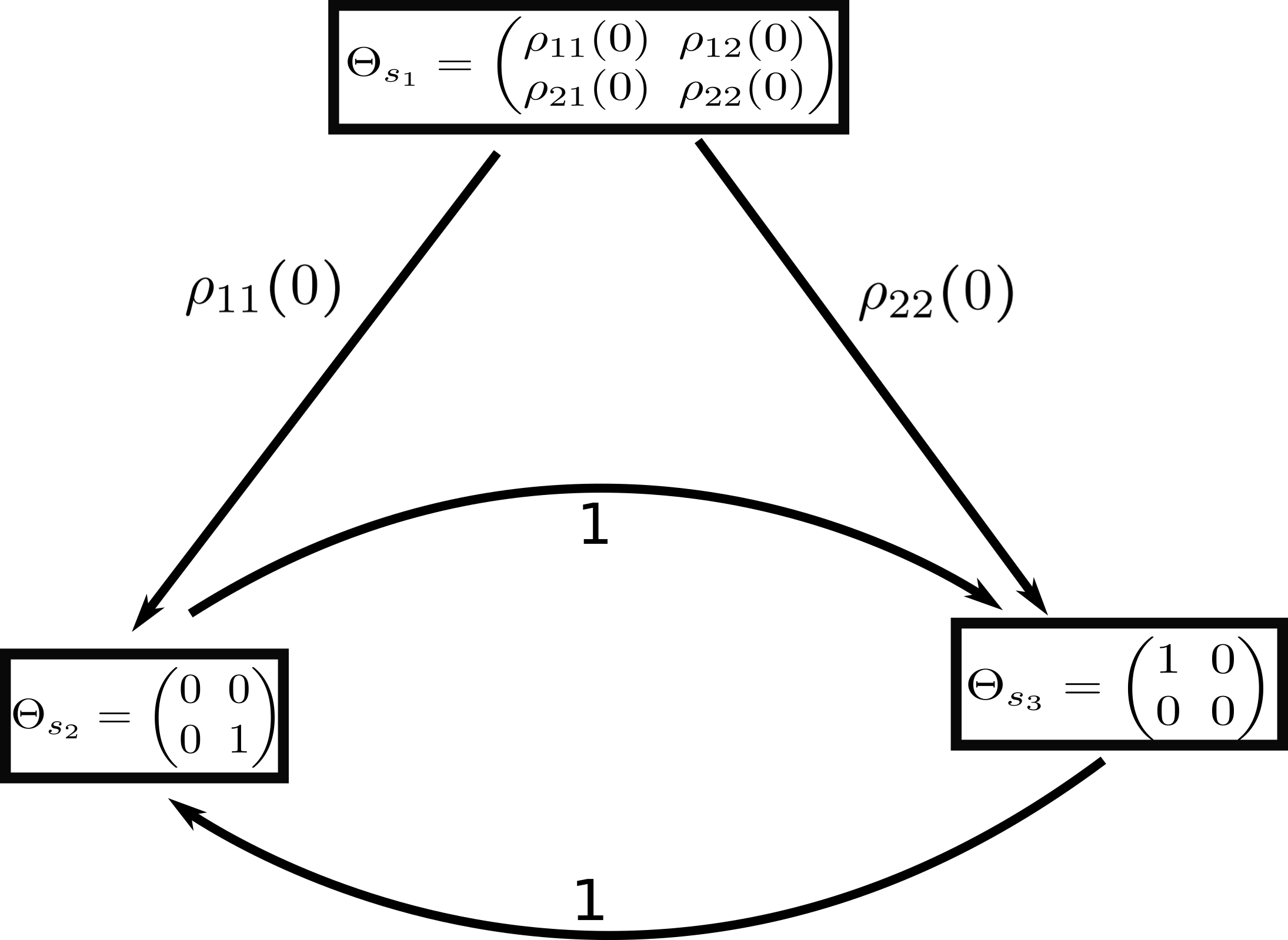}
\caption{State transition network for example a). }
\label{Example_a_StateTransitionNetwork} 
\end{center}
\end{figure}

\subsubsection[Example b)]{Example b): $H \propto \Id$,  $V_1 = \begin{pmatrix}
 1 & 0 \\
 0 & 0
 \end{pmatrix}$,  $\Lambda = \begin{pmatrix}
 \g_1 & 0 \\
 0 & 0
 \end{pmatrix}$.  }
 
The set of possible states appearing in the Markov chain can be computed to: 

\begin{equation}
\begin{aligned}
\Omega_{\brho_0}
=
\left\{
\Theta_{s_1} = \begin{pmatrix}
 \brho_{11}(0) & \brho_{12}(0) \\
\brho_{21}(0)& \brho_{22}(0)
 \end{pmatrix}, 
\Theta_{s_2} = 
\begin{pmatrix}
0 & 0 \\
0 & 1
\end{pmatrix},
\Theta_{s_3} = 
\begin{pmatrix}
1 & 0 \\
0 & 0
\end{pmatrix}
\right\}. 
\end{aligned}
\end{equation}

In this case, $\Theta_{s_1}=\brho_0$ is a possible trapping state, since without it the Markov chain is incomplete, as the column sum does not equal one. 

\begin{figure}[H]
\begin{center}
\begin{subfigure}{0.49\textwidth}
\subcaption{Incomplete Markov chain: 
$
Q =
\begin{pmatrix}
0 & 0  \\
\brho_{11}(0) & 1
 \end{pmatrix}. 
$
}
\hspace*{20mm}
\includegraphics[width=0.5\columnwidth]{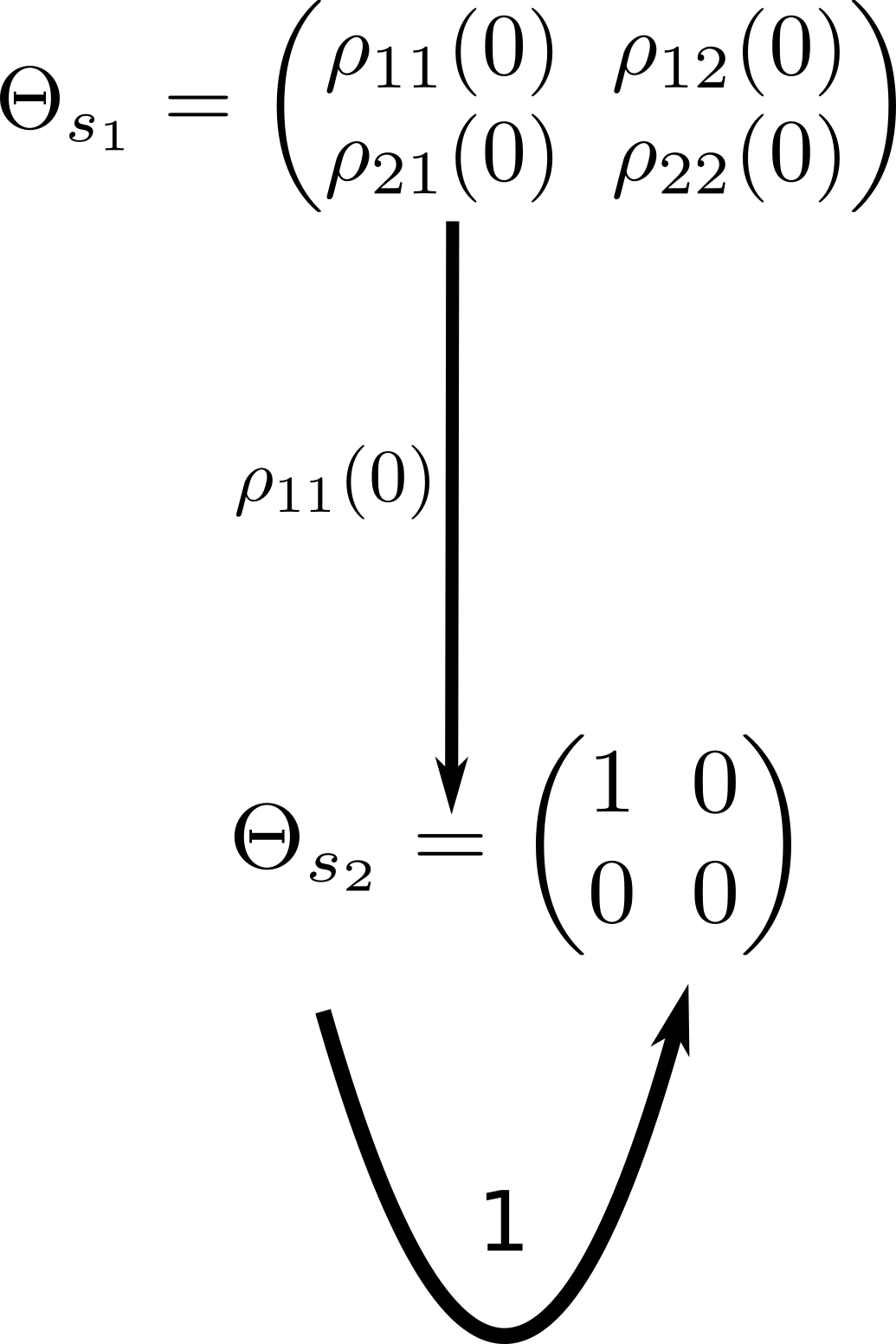} 
\label{Example_b_StateTransitionNetwork_1}
\end{subfigure}\begin{subfigure}{0.49\textwidth}
\subcaption{Complete Markov chain: 
$
Q =
\begin{pmatrix}
0 & 0 & 0 \\
\brho_{11}(0) & 1 & 0 \\
\brho_{22}(0) & 0 & 1
 \end{pmatrix}. 
$}
\includegraphics[width=0.8\columnwidth]{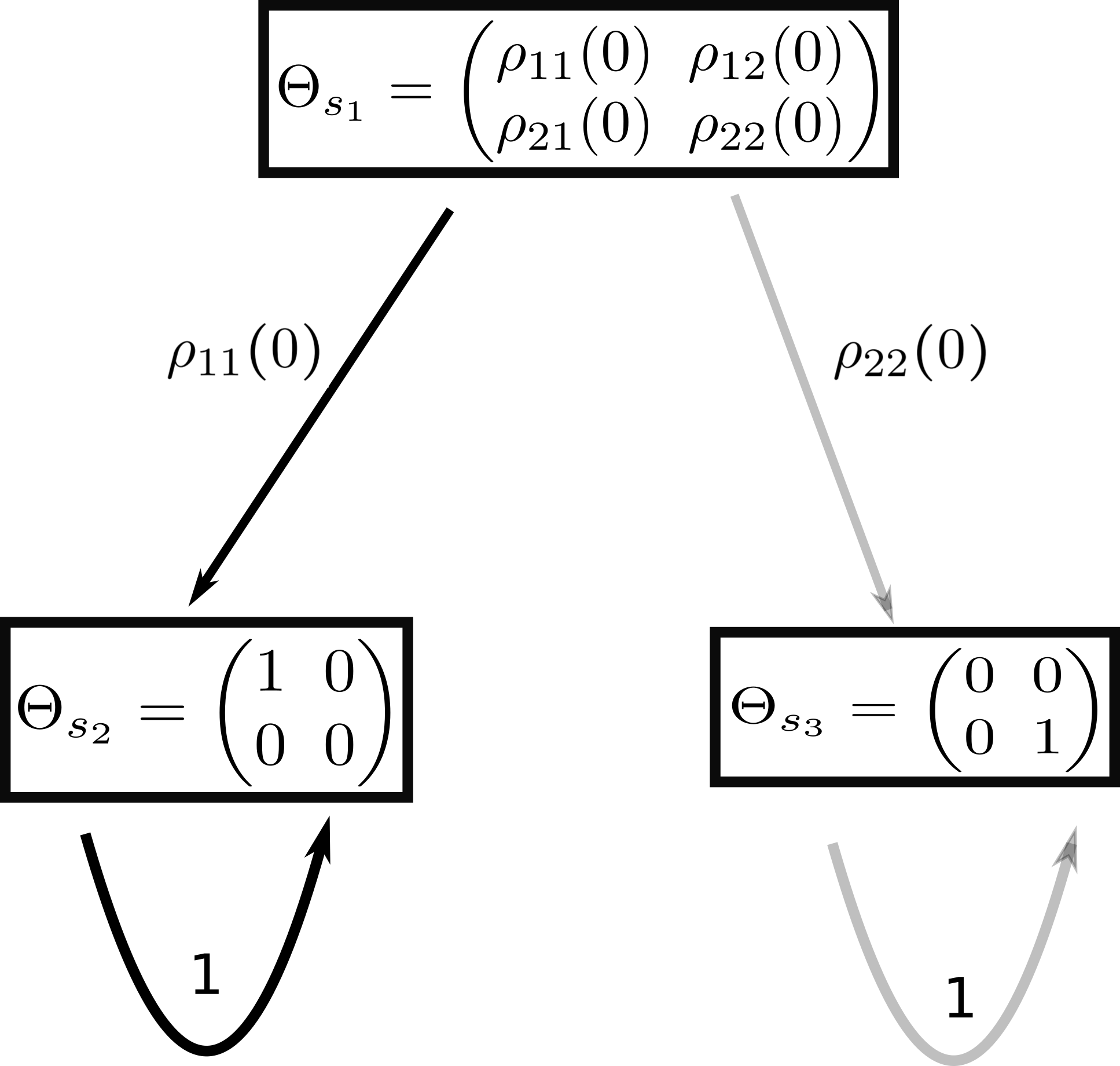} 
\label{Example_b_StateTransitionNetwork_2}
\end{subfigure}
\caption{State transition network for example b).}
\label{Example_b_StateTransitionNetwork} 
\end{center}
\end{figure}

\subsubsection[Example c)]{Example c): $H \propto \Id$,  $V_1 = \begin{pmatrix}
 0 & 0 \\
1 & 0
 \end{pmatrix}$, 
 $\Lambda = \begin{pmatrix}
 \g_1 & 0 \\
 0 & 0
 \end{pmatrix}$. 
 }

The set of possible states appearing in the Markov chain can be computed to: 

\begin{equation}
\begin{aligned}
\Omega_{\brho_0}
=
\left\{
\Theta_{s_1} = \begin{pmatrix}
 \brho_{11}(0) & \brho_{12}(0) \\
\brho_{21}(0)& \brho_{22}(0)
 \end{pmatrix}, 
\Theta_{s_2}
 =
 \begin{pmatrix}
 0 & 0 \\
 0 & 1
 \end{pmatrix} 
 = \Theta_{s_3}
 \right\}. 
\end{aligned}
\end{equation}

Both $\Theta_{s_1}=\brho_0$ and $\Theta_{s_2}$ are possible trapping states, even though there is just one minimal absorbing set, namely 
$$B = \{\Theta_{s_2}\} = 
\left\{ \lim\limits_{T\to \infty} \frac{I_{\Theta_{s_1}}(T)}{T}\right\}  = \left\{
\begin{pmatrix}
0 & 0 \\ 0 & 1
\end{pmatrix}
\right\} = \left\{ \lim\limits_{T\to \infty} \frac{I_{\Theta_{s_2}}(T)}{T}\right\}. $$

\begin{figure}[H]
\begin{center}
\begin{subfigure}{0.49\textwidth}
\vspace*{-9mm}
\subcaption{Incomplete Markov chain: 
$
Q =
\begin{pmatrix}
0 & 0  \\
\brho_{11}(0) & 0
 \end{pmatrix}. 
$
}
\hspace*{15mm}
\includegraphics[width=0.4\columnwidth]{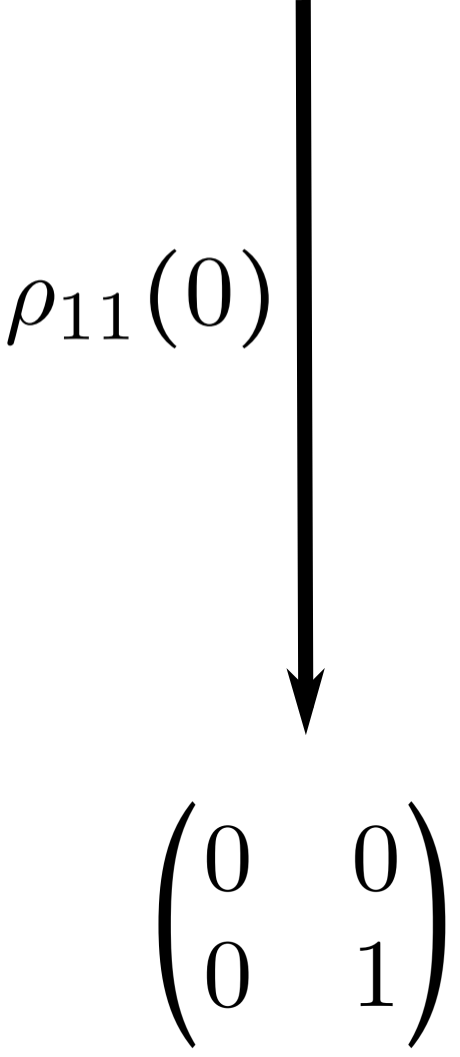} 
\label{Example_c_StateTransitionNetwork_1}
\end{subfigure}
\begin{subfigure}{0.49\textwidth}
\subcaption{Complete Markov chain: 
$
Q =
\begin{pmatrix}
0 & 0  \\
1 &  0 
\end{pmatrix}. 
$}
\hspace*{17mm}
\includegraphics[width=0.6\columnwidth]{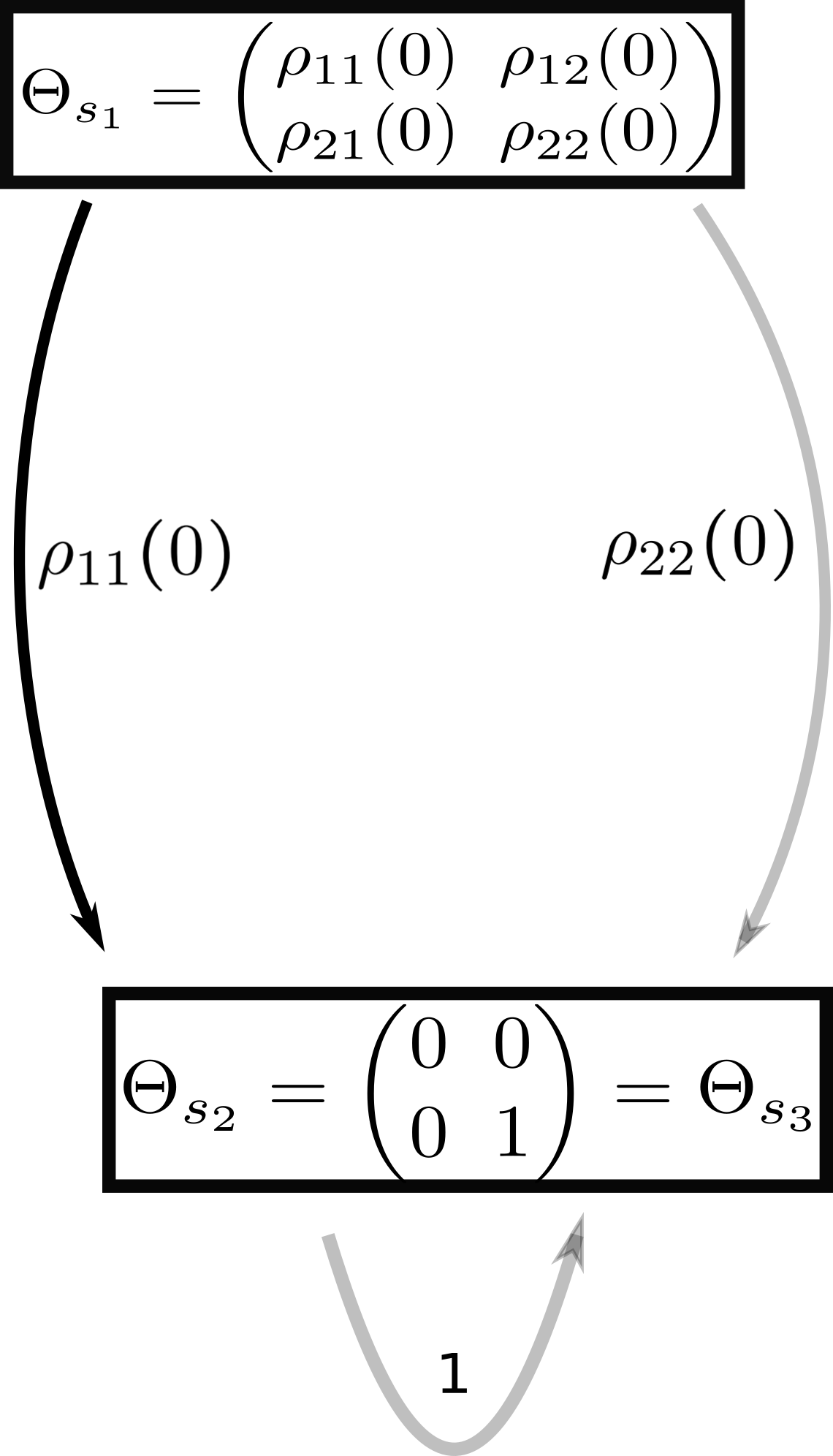} 
\label{Example_c_StateTransitionNetwork_2}
\end{subfigure}
\caption{State transition network for example c).}
\label{StateTransitionNetwork_1_2} 
\end{center}
\end{figure}

\subsection{Evaluating the stationary state $\Theta_B$ of the Lindbladian for the minimal absorbing set $B$}
 \label{Evaluating_Theta_BForAGeneralConditionalHamiltonian_H_c}

We now evaluate the 
stationary states $\Theta_B$ occurring in equation \eqref{Theta_B}.

We write the conditional Hamiltonian $H_c$ (see equation \eqref{Def_H_c}) as a direct sum of Jordan blocks

\begin{equation}
\begin{aligned}
H_c 
&=
\bigoplus_{m=1}^M J_m 
=
\begin{pmatrix}
J_1 & & \\
 & \ddots & \\
 & & J_M
\end{pmatrix}, 
\text{  with  } \\
J_m 
&=
\begin{pmatrix}
\delta_m & 1 &  \\
 & \ddots &  \ddots \\ 
 & & \delta_m & 1 \\
 & & & \delta_m
\end{pmatrix} \in \C^{s_m \times s_m} 
\end{aligned}
\end{equation}
being a Jordan block of size $s_m\in \N$ and $\sum\limits_{m=1}^M s_m = N$.

We discuss separately the two cases of a finite and an infinite number of quantum jumps.

\subsubsection{Infinitely many quantum jumps}\label{Evaluate_I_s_of_tau_s}

When the quantum trajectory contains infinitely many quantum jumps, 
we need to evaluate the expression $\overline{I_s(\tau_s)}$ defined in equation \eqref{LawOfLargeNumbers_Applied}.

When $\Theta_s$ is not a possible trapping state, we can compute the (conditional) time evolution operator $\e^{-i \, H_c \, t}$ (for details see appendix \ref{AuxiliaryCalculations}) and get

\begin{equation}\label{ExpressionOf_I_s_of_tau_s}
\begin{aligned}
\overline{I_s(\tau_s)} 
&=
\int_0^\infty 
\underbrace{
\hspace*{3mm}
f(\tau \,|\, \Theta_s)
\hspace*{3mm}
}_{
- \frac{\d}{\d \tau} \Tr[U_\tau(\Theta_s)]
}
\,
\underbrace{
\hspace*{3mm}
I_{\Theta_s}(\tau)
\hspace*{3mm}
}_{
\int_0^\tau
\frac{U_t(\Theta_s)}{\Tr[\dots]} \, \d t
}
\d \tau = \\
&\xlongequal[\text{integration}]{\text{partial}}
-  \Tr[U_\tau(\Theta_s)] 
\cdot
\int_0^\tau
\frac{U_t(\Theta_s)}{\Tr[\dots]} \, \d t \;
\mathlarger{\Bigr|}_{\tau=0}^{\tau=\infty}
+
\int_0^\infty
\Tr[U_\tau(\Theta_s)] \;  \frac{U_\tau(\Theta_s)}{\Tr[U_\tau(\Theta_s)]}
\d \tau= \\
&=
\underbrace{
-\lim\limits_{\tau \to \infty} \, 
 \Tr[U_\tau(\Theta_s)] 
\cdot
\int_0^\tau
\frac{U_t(\Theta_s)}{\Tr[\dots]} \, \d t \;
}_{
0
}
+
\int_0^\infty U_\tau(\Theta_s) \d \tau = \\
&\xlongequal{\ref{AuxiliaryCalculations}}
\int_0^\infty U_\tau(\Theta_s) \d \tau
\xlongequal{\text{Def } U_\tau()}
\int_0^\infty \, 
\e^{-i \, H_c \, \tau} \, \Theta_s \, (\e^{-i \, H_c \, \tau} )^\dagger \, \d \,  \tau = \\ 
&=
\int_0^\infty \, 
\underbrace{
\begin{pmatrix}
\e^{-i \, J_1 \, \tau} & & 0 \\
& \ddots & \\
0 & & \e^{-i \, J_M \, \tau}
\end{pmatrix} \, 
\begin{pmatrix}
\hat{\bTheta}^{(1,1)} & \dots & \hat{\bTheta}^{(1,M)} \\
\vdots & & \vdots \\
\hat{\bTheta}^{(M,1)} & \dots & \hat{\bTheta}^{(M,M)} 
\end{pmatrix} \, 
\begin{pmatrix}
\e^{i \, J_1^T \, \tau} & & 0 \\
& \ddots & \\
0 & & \e^{i \, J_M^T \, \tau}
\end{pmatrix}
}_{ 
\begin{pmatrix}
\e^{-i \, J_1 \, \tau} \, \hat{\bTheta}^{(1,1)} \, \e^{i \, J_1^T \, \tau} 
& \dots &
\e^{-i \, J_1 \, \tau} \, \hat{\bTheta}^{(1,M)} \, \e^{i \, J_M^T \, \tau} \\ 
\vdots & & \vdots \\ 
\e^{-i \, J_M \, \tau} \, \hat{\bTheta}^{(M,1)}\, \e^{i \, J_1^T \, \tau} 
& \dots &
\e^{-i \, J_M \, \tau} \, \hat{\bTheta}^{(M,M)} \, \e^{i \, J_M^T \, \tau} \\ 
\end{pmatrix}
}
\, \d \,  \tau = \\ \\ 
&=
\begin{pmatrix}
\int_0^\infty \e^{-i \, J_1 \, \tau} \, \hat{\bTheta}^{(1,1)} \, \e^{i \, J_1^T \, \tau} \, \d \tau
& \dots & 
\int_0^\infty\e^{-i \, J_1 \, \tau} \, \hat{\bTheta}^{(1,M)} \, \e^{i \, J_M^T \, \tau} \, \d \tau \\ 
\vdots & & \vdots \\ 
\int_0^\infty\e^{-i \, J_M \, \tau} \, \hat{\bTheta}^{(M,1)}\, \e^{i \, J_1^T \, \tau} \, \d \tau
& \dots & 
\int_0^\infty \, \e^{-i \, J_M \, \tau} \, \hat{\bTheta}^{(M,M)} \, \e^{i \, J_Y^T \, \tau} \, \d \tau  
\end{pmatrix}. 
\end{aligned}
\end{equation}

For the component $\Bigl( \sum_{\mu=1}^{m-1} s_\mu + j, \sum_{\nu=1}^{n-1} s_\nu + k \Bigr)$, we get with $\delta_x = R_x + i \, I_x$: 
\begin{equation}\label{I_s_of_tau_s_evaluated}
\begin{aligned}
\Bigl(\overline{I_s(\tau_s)}\Bigr)_{\Bigl( \sum_{\mu=1}^{m-1} s_\mu + j, \sum_{\nu=1}^{n-1} s_\nu + k \Bigr)}
&= \left( \int_0^\infty \, \e^{-i \, J_m \, \tau} \, \hat{\bTheta}^{(m,n)} \, \e^{i \, J_n^T \, \tau} \, \d \tau  \right)_{j \, k} \\
&\xlongequal{(*)}
\begin{cases}
\sum_{\alpha=j}^{s_m} \, \sum_{\beta=k}^{s_n} \, 
\colvec{2}{\alpha-j+\beta-k}{\alpha-j} \, 
\frac{(-1)^{\alpha-j+\beta-k} \, (\hat{\bTheta}^{(m,n)})_{jk} }{[i \, (R_n - R_m) + I_n + I_m]^{\alpha-j+\beta-k}}, &\text{ if  } I_n\neq 0 \neq I_m \\
0 &\text{ else.  }
\end{cases}
\end{aligned}
\end{equation}

In step (*) in the last line
we used the fact that since $\Theta_s$ is not a possible trapping state, a diagonal block of $\Theta_s$ must vanish whenever the corresponding eigenvalue of $\Lambda$ vanishes $\left(\lambda_m = 0 \Longrightarrow \hat{\bTheta}_s^{(m,m)}=\bzero^{s_m \times s_m}\right)$, \text{ see appendix \ref{AuxiliaryCalculations}},  and an inequality that is valid for all density matrices: $|(\Theta_s)_{jk}| \, \leq (\Theta_s)_{jj} \cdot (\Theta_s)_{kk}$.

An important special case is when both Jordan blocks are of size one, that is $s_m=1=s_n$.  
In that case we have

\begin{equation}
\begin{aligned}
\left(
\int_0^\infty \, 
\underbrace{
\e^{-i \, J_m \, t}
}_{
\e^{-i \, \delta_m \, t}
}\, \hat{\bTheta}^{(m,n)} \, 
\underbrace{
\e^{i \, J_n^T \, t}
}_{
\e^{i \, \delta_n^* \, t}
}
 \, \d \, t
\right)_{11}
&=
\begin{cases}
\frac{\Bigl(\hat{\bTheta}^{(m,n)}\Bigr)_{11} }{[i \, (R_n - R_m) + I_n + I_m]}, &\text{ if } I_n \neq 0 \neq I_m \\
0, &\text{ else.  }
\end{cases}
\end{aligned}
\end{equation}

When the conditional Hamiltonian $H_c$ is diagonalizable, then all Jordan block are of size one, the block matrices $\hat{\bTheta}^{(m,n)} \in \C$ are scalars, and expression \eqref{I_s_of_tau_s_evaluated} becomes: 

\begin{equation}
\begin{aligned}
\Bigl(\overline{I_s(\tau_s)}\Bigr)_{m\, n}
=
\frac{\Bigl(\hat{\bTheta}^{(m,n)}\Bigr)_{11} }{i\,(\delta_n^* - \delta_m)}
=
\begin{cases}
\frac{(\Theta_s)_{m\, n} }{[i \, (R_n - R_m) + I_n + I_m]}, &\text{ if } I_n \neq 0 \neq I_m \\
0, &\text{ else.  }
\end{cases} 
\end{aligned}
\end{equation}

\subsubsection{ Finitely many quantum jumps }\label{Theta_B_FinitelyManyQuantumJumps}

When the number of jumps is finite, we have to evaluate $\Theta_B = \left\langle \frac{U_t(\Theta_\text{trap})}{\Tr[]} \right\rangle_{t\geq 0}$, with $\Theta_\text{trap}$ being a possible trapping state.

By definition of a possible trapping state, we know that $\Tr[U_t(\Theta_\text{trap})]$ converges to a positive value: $\Tr[U_t(\Theta_\text{trap})] \xlongrightarrow{t \to \infty} q \in (0,1]$. 
Then we have

\begin{equation}\label{Auxilliary_2_a}
\begin{aligned}
\frac{\Bigl(U_t(\Theta_\text{trap})\Bigr)_{\Bigl(\sum\limits_{\mu=1}^{m-1} s_\mu + j, \, \sum\limits_{\nu=1}^{n-1} s_\nu + k\Bigr)}}{\Tr[U_t(\Theta_\text{trap})]}
=
\frac{
\sum\limits_{\alpha=j}^{s_m} \,
\sum\limits_{\beta=k}^{s_n} \,
\e^{i \, (R_n - R_m) \, t} \, \e^{(I_n + I_m) \, t}
\, 
\frac{t^{(\alpha-j)}}{(\alpha-j)!} \, 
\frac{t^{(\beta-k)}}{(\beta-k)!} 
\left(\hat{\bTheta}^{(m,n)} \right)_{\alpha, \beta}
}{
\underbrace{
\Tr[U_t(\Theta_\text{trap})]
}_{
\xlongrightarrow{t \to \infty}
 q 
}
}
\xlongrightarrow[\text{if  } I_n < 0 \text{ or } I_m < 0]{t \to \infty}
0
\end{aligned}
\end{equation}

and 

\begin{equation}\label{Auxilliary_2_b}
\begin{aligned}
\lim\limits_{T \to \infty}
\frac{1}{T}
\int_0^T \e^{i \, \Delta E \, t} \, \d \, t
= 
\left.
\begin{cases}
1 &\text{ ,if } \Delta E = 0 \\
\lim\limits_{T \to \infty}
\frac{\e^{i \, \Delta E \, t}  - 1}{i \, \Delta E \, T} = 0 &\text{, else } 
\end{cases}
\right\}
 \hspace*{4mm} = 
\delta_{\Delta \, E, 0}. 
\end{aligned}
\end{equation}

When we combine the equations \eqref{Auxilliary_2_a} and \eqref{Auxilliary_2_b}, we get 

\begin{equation}
\begin{aligned}
\lim\limits_{T \to \infty}
\frac{1}{T} \int_0^T 
\left(
\frac{U_t(\Theta_\text{trap})}{\Tr[]}
\right)_{\Bigl(\sum\limits_{\mu=1}^{m-1} s_\mu + j, \, \sum\limits_{\nu=1}^{n-1} s_\nu + k\Bigr)}
\, \d \, t
=
\delta_{I_n, 0} \, 
\delta_{I_m, 0} \, 
\delta_{R_n, R_m} 
\, \cdot \, 
\frac{ (\bTheta^{(m,n)})_{j, \, k} }
{\sum\limits_{\mu \in \{1, \dots, M\,:\, I_\mu = 0\}} \left(\bTheta^{m,n}\right)_{\mu, \mu}}.
\end{aligned}
\end{equation}

\subsection{The stationary solution of the Lindblad equation}\label{TheStationarySolutionOfTheLindbladEquation}

Now that we have obtained all building blocks for the stationary solution, let us put them together. Our starting point was the quantum jump unravelling (section \ref{secunravel}) and the fact that the time and ensemble average can be exchanged (section \ref{ChangingLimits_TimeAndEnsembleAverage})
, due to the fact that the time average of a single trajectory exists. 
We assumed that the space of density matrices occurring immediately after jumps is finite  $(|\s|< \infty)$, so that a stationary solution 
of the corresponding Markov chain
always exists.
We evaluated the time average $\Theta_B := \Theta_{\brho_0}^\infty(\omega) = \langle\Theta_{\brho_0}(\omega, t) \rangle_{t \geq 0})$ of a single trajectory $\omega \in \Unravel$
(see equation \eqref{Theta_B} in section \ref{TimeAverageOfASingleQuantumTtrajectory}), and we showed  that it depends only on the minimal absorbing set $B\subseteq \Omega_{\brho_0}$ the quantum trajectory is eventually captured in (section \ref{DiscreteTimeMarkovChainsWithFiniteStateSpace}). We therefore separated the integral over the different unravellings into a sum over the minimal absorbing sets in the network, where the integrand  $\Theta_B$ is constant over the set $\Unravel(B)$ of quantum trajectories that all land in the associated minimal absorbing set $B\subseteq \Omega_{\brho_0}$.  
Mathematically, all this gives the following steps for obtaining the formula for the limiting density matrix $\brho^\infty(\brho_0)$, depending on the initial state $\brho_0$:

%
\begin{equation}\label{Computing_rho_infty_of_rho_0}
\begin{aligned}
\brho^\infty(\brho_0)
&=
\lim\limits_{T \to \infty}
\frac{1}{T}
\int_0^T 
\brho(t) \, \d \,t
=
\langle
\brho_{\brho_0}(t)
\rangle_{t\geq \, 0}
=
\langle
\langle
\Theta_{\brho_0}(t, \omega)
\rangle_{\omega \in U}
\rangle_{t\geq \, 0} =  \\
%
&\xlongequal{\eqref{ChangingLimits_TimeAndEnsembleAverage}}
\langle
\;
\underbrace{
\langle
\Theta_{\brho_0}(t, \omega)
\rangle_{t\geq \, 0}
}_{
\Theta_{\brho_0}^{\infty}(\omega)
}\;
\rangle_{\omega \in U}
=
\int_{\Unravel = \dot{\bigcup}_{i=1}^n \Unravel(B_i)} 
\Theta_{\brho_0}^{\infty}(\omega) \, \d \,  \Prob_{\brho_0}(\omega) =\\
%
&\xlongequal{|\s|<\infty}
\sum\limits_{i=1}^n
\int_{\Unravel(B_i)} 
\underbrace{
\Theta_{\brho_0}^{\infty}(\omega) 
}_{
\Theta_{B_i}
}
\, \d \,\Prob_{\brho_0}(\omega)
=
\sum\limits_{i=1}^n
\,
\Prob\bigl(\Unravel(B_i)\,|\, \brho_0\bigr) \, \Theta_{B_i}. 
\end{aligned}
\end{equation}

By mapping the quantum trajectories onto  discrete-time Markov chains (section \ref{ConstructingTheMarkovChainForQuantumTrajectories}), we were able to compute the probability $\Prob\bigl(\Unravel(B_i)\,|\, \brho_0\bigr)$ of landing in the minimal absorbing set $B_i$ as the coefficient $\mu_{B_i}$ in an expansion of  the initial probability distribution vector $\bq_{\brho_0}$ into an ordered basis of generalized eigenvectors of the transition matrix $Q$ (see section \ref{DiscreteTimeMarkovChainsWithFiniteStateSpace},

\begin{equation}
\begin{aligned}
\bq_{\brho_0}
=
\sum\limits_{i=1}^{|\s|} \mu_i \, \bh^{i}
=:
\sum\limits_{i=1}^{n} \mu_{B_i} \, \bq^\infty(\bq_{\brho_0} \in B_i)
+
\sum\limits_{i=n+1}^{|\s|} \mu_i \, \bh^{i}. 
\end{aligned}
\end{equation}

 The first $n$ generalized eigenvectors $\bq^\infty(\bq_{\brho_0} \in B_i)$ for $i \in \{1, \dots n\}$  are the actual eigenvectors of $Q$ to the eigenvalue $\lambda=1$ (see equation \eqref{FormulaFor_p_infty} in Appendix \ref{FormOfStationarySolutions}).
  The second term contains the remaining generalized eigenvectors $\bh^{i}$, and it declines toward zero under repeated application of the transition matrix $Q$.

An explicit expression for $\Theta_B$ was given in section \ref{Evaluating_Theta_BForAGeneralConditionalHamiltonian_H_c}. It is 
the stationary solution of the Lindblad equation that lies in the minimal absorbing set $B\subseteq \Omega_{\brho_0}$. 

\subsection{The classical case}\label{TheClassicalCase}

The classical case can be recovered when the Hamiltonian vanishes ($H=0$) and $\Lambda = \beta \, \Id$ is proportional to the identity operator for some real number $\beta\in \R_{\geq \, 0}$. In this case, there are no possible trapping states, and we get for the stationary solution of a minimal absorbing set $B \subseteq \Omega_{\brho_0}$ and for the stationary solution of equation \eqref{Lindbladian}

\begin{equation}
\begin{aligned}
\Theta_B
&=
\frac{\sum\limits_{s \in \s} \, q(\Theta_s \,|\, B )  \, 
\overbrace{
\overline{I_s(\tau_s)}
}^{\Theta_s/\beta}
}{
\sum\limits_{s \in \s} \, q(\Theta_s \,|\, B )  \,  \frac{1}{\beta} \, \underbrace{
\Tr[\Theta_s]
}_{
1
}
}
=
\frac{\sum\limits_{s \in \s} \, q(\Theta_s \,|\, B )  \, \Theta_s \, \frac{1}{\beta}}{ \, \frac{1}{\beta}}
=
\sum\limits_{s \in \s} \, q(\Theta_s \,|\, B )  \, \Theta_s \\
\brho^\infty(\brho_0)
&=
\sum\limits_{i=1}^n
\,
\Prob_{\brho_0}\bigl(\Unravel(B_i)\bigr) \, \Theta_{B_i}
\,
=
\sum\limits_{s \in \s} \, 
\underbrace{
\left(
\sum\limits_{i=1}^n \,\Prob_{\brho_0}\bigl(\Unravel(B_i)\bigr)  \,  q(\Theta_s \,|\, B_i)
\right)
}_{
q(\Theta_s)
}\, \Theta_s
=
\sum\limits_{s \in \s} \, q(\Theta_s) \,  \Theta_s. 
\end{aligned}
\end{equation}

\section{Discussion and Conclusion}

We have shown that the stationary states of the Lindblad equation \eqref{Lindbladian} can be determined by using the concept of a \textit{quantum jump unravelling}. The final formula is given by
\begin{equation}\label{Computing_rho_infty_of_rho_0_2}
\begin{aligned}
\brho^\infty(\brho_0)
=
\sum\limits_{i=1}^n
\,
\Prob\bigl(\Unravel(B_i)\,|\, \brho_0\bigr) \, \Theta_{B_i}, 
\end{aligned}
\end{equation}
where the sum goes over all minimal absorbing set of the discrete-time Markov chain associated with the quantum jump process, whose states $\{\Theta_s \,:\, s \in \s \}$ and transition probabilities  $\Prob(\Theta_{s_1} \to \Theta_{s_1})$ are given in section \ref{TheStatesOfTheMarkovChain} and \ref{TheTransitionProbabilitiesOfTheMarkovChain}, respectively. 

The factor $\Theta_{B_i} = \lim\limits_{T \to \infty} \frac{1}{T} \int_0^T \, \Theta(t, \omega) \d \, t$ denotes the time average of all quantum trajectories $\omega\in \Unravel$ which are eventually captured by the minimal absorbing set $B_i \subseteq \Omega_{\brho_0}$. Explicit expressions for $\Theta_{B_i}$ are derived in section \ref{Evaluating_Theta_BForAGeneralConditionalHamiltonian_H_c}, depending on whether the number of quantum jumps is finite or infinite. 
$\Prob\bigl(\Unravel(B_i)\,|\, \brho_0\bigr) $ denotes the probability for a quantum trajectory being captured by the minimal absorbing set $B_i$, provided the initial state is given by $\brho_0$. It can be computed from the coefficients $\mu_{B_i}$ of the vector $\bq_{\brho_0}$ (representing the initial state $\brho_0 \in \Omega_{\brho_0}$) when expressing it in a basis of generalized eigenvectors of the transition matrix $Q$ of the discrete-time Markov chain, which is defined in section \ref{TheTransitionProbabilitiesOfTheMarkovChain}.
 When there is no change of the states between jumps, the classical case is recovered, see section \ref{TheClassicalCase}. 

In the case of infinitely many states, such a methods is in general no longer applicable:
Although the existence of stationary density matrices $\brho^\infty \in \C^{N\times N}$ for the \textit{Lindbladian} is guaranteed (see \cite{kuemmerer2004pathwise}), the existence of stationary vectors $\bq^\infty \in \R^{|\s|}$ for the corresponding \textit{Markov chain} is not, as shown in appendix.  \ref{TheNeedForAFiniteStateSpace}. This limitation can only be dissolved if an analogous expression to formula \eqref{FormulaFor_q_infty}, but for special kinds of Markov chains with infinite state space is known. One possible condition could be to require the Markov chain to be \textit{positive recurrent}, in which case a stationary solution exists \cite{privault2013understanding,douc2018markov,bremaud2013markov}. An interesting question for further research would be which conditions the Lindblad operators must have for the states of the quantum trajectory to be positive recurrent or whether this is a necessary condition for the Lindblad operators.

However, it is possible to compute the stationary density matrix $\brho^\infty$ when the sequence of states in the quantum trajectory, together with the corresponding average waiting times, converges: $(\Theta_k, \overline{\tau_k}) \xrightarrow{k \to \infty} (\Theta_\infty, \overline{\tau_\infty})$. In this case, we can evaluate the time average of a quantum trajectory to
\begin{equation}\label{TimeAverageOfQuantumTrajectoryWhenBHasInfinitelyManyStates}
\begin{aligned}
\langle\Theta(t, \omega) \rangle_{t \geq 0}
&=
\lim\limits_{T \to \infty}
\frac{
\frac{1}{J(T)} \, \sum\limits_{k=0}^{J(T)-1} I_{\Theta_k}(\tau_k)
}{\Tr[]}
=
\frac{1}{\overline{\tau_\infty}}
\int_0^{\overline{\tau_\infty}}
\frac{U_t(\Theta_\infty)}{\Tr[]} \, \d t. 
\end{aligned}
\end{equation}
This means that while we can no longer use the formula for $\Theta_B$ given by equation \eqref{Theta_B} when the minimal absorbing set $B$ contains infinitely many states, we can can use equation \eqref{Computing_rho_infty_of_rho_0_2} when the \textit{number} of minimal absorbing sets is finite. We only have to substitute the expression of equation \eqref{TimeAverageOfQuantumTrajectoryWhenBHasInfinitelyManyStates} for $\Theta_{B_i}$. Unfortunately, this is not quite the same as giving an explicit recipe for obtaining the steady state, as was possible in the case with finitely many states. But it might still be possible to obtain the stationary density matrices this way for many more cases that are physically relevant.


\appendix

\section{Appendix}\label{Appendix}

\subsection{The time average and the ensemble average \cite{kuemmerer2004pathwise}}\label{ChangingLimits_TimeAndEnsembleAverage}

For almost all $\omega \in \Unravel $, we have

\begin{align*}
\lim\limits_{T \to \infty} \frac{1}{T} \int_0^T d t \;  \Theta_{\brho_0} (t, \omega) =  \Theta_{\brho_0}^\infty (\omega). 
\end{align*}
Here $\Theta_{\brho_0}^\infty$ is a random variable whose expectation value is the time average of $U(t) \, \brho_0$,
\begin{align*}
\langle\langle  \Theta_{\brho_0} (t, \omega) \rangle_{t\geq 0 }\rangle_{\omega \in \Unravel} = 
\langle\langle  \Theta_{\brho_0} (t, \omega) \rangle_{\omega \in \Unravel}\rangle_{t\geq 0 }. 
\end{align*}

So, in this sense, the time average commutes with the ensemble average.  

\subsection{Auxiliary calculations for section \ref{Evaluate_I_s_of_tau_s}}\label{AuxiliaryCalculations}

\subsubsection{The form of the (conditional) time evolution operator}

With respect to a suitable basis, we can write the conditional Hamiltonian in the Jordan normal form, that is 

\begin{equation} \label{JordanNormalFormOf_Hc}
\begin{aligned}
H_c 
&=
\bigoplus_{m=1}^M J_m 
=
\begin{pmatrix}
J_1 & & \\
 & \ddots & \\
 & & J_M
\end{pmatrix}, 
\text{  with  } \\
J_m 
&=
\begin{pmatrix}
\delta_m & 1 &  \\
 & \ddots &  \ddots \\ 
 & & \delta_m & 1 \\
 & & & \delta_m
\end{pmatrix} \in \C^{s_m \times s_m}, \text{  with } \sum\limits_{m=1}^M s_m = N. 
\end{aligned}
\end{equation}

The eigenvalues of $H_c$ are then given by $\{\delta_1,\dots, \delta_M\}$

This makes it easier to compute matrix exponentials, in particular the conditional time evolution operator

\begin{equation} \label{ConditionalTimeEvolutionOperator}
\begin{aligned}
\e^{-i \, H_c \, t}
&=
\begin{pmatrix}
\e^{-i \, J_1 \, t} & & 0 \\
& \ddots & \\
0 & & \e^{-i \, J_M \, t}
\end{pmatrix}
=
\bigoplus_{m=1}^M \e^{-i \, J_m \, t }
=
\bigoplus_{m=1}^M \, 
\e^{-i \, \delta_m \, t}
\begin{pmatrix}
1 & \frac{t^1}{1!} & \dots &  & \frac{t^{(s_m-1)}}{(s_m-1)!} \\
& 1 & \frac{t^1}{1!} & & \frac{t^{(s_m-2)}}{(s_m-2)!} \\
&  & \ddots & \ddots & \vdots \\
& & & \ddots & \frac{t^1}{1!}  \\ 
& & &  & 1
\end{pmatrix}
\end{aligned}
\end{equation}

When applying the conditional time evolution to a density matrix $\Theta_s$, we get 

\begin{equation} \label{ContitionalTimeEvolutionOfTheta_s}
\begin{aligned}
U_\tau(\Theta_s)
&=
\begin{pmatrix}
\e^{-i \, J_1 \, \tau} & & 0 \\
& \ddots & \\
0 & & \e^{-i \, J_M \, \tau}
\end{pmatrix} \, 
\begin{pmatrix}
\hat{\bTheta}^{(1,1)} & \dots & \hat{\bTheta}^{(1,M)} \\
\vdots & & \vdots \\
\hat{\bTheta}^{(M,1)} & \dots & \hat{\bTheta}^{(M,M)} 
\end{pmatrix} \, 
\begin{pmatrix}
\e^{i \, J_1^T \, \tau} & & 0 \\
& \ddots & \\
0 & & \e^{i \, J_M^T \, \tau}
\end{pmatrix} = \\
&=
\begin{pmatrix}
\e^{-i \, J_1 \, \tau} \, \hat{\bTheta}^{(1,1)} \, \e^{i \, J_1^T \, \tau} 
& \dots &
\e^{-i \, J_1 \, \tau} \, \hat{\bTheta}^{(1,M)} \, \e^{i \, J_M^T \, \tau} \\ 
\vdots & & \vdots \\ 
\e^{-i \, J_M \, \tau} \, \hat{\bTheta}^{(M,1)}\, \e^{i \, J_1^T \, \tau} 
& \dots &
\e^{-i \, J_M \, \tau} \, \hat{\bTheta}^{(M,M)} \, \e^{i \, J_M^T \, \tau}. 
\end{pmatrix}
\end{aligned}
\end{equation}

When we look at the components $\Bigl( \sum\limits_{\mu=1}^{m-1} s_\mu + j, \sum\limits_{\nu=1}^{n-1}s_\nu + k \Bigr)$ of equation \eqref{ContitionalTimeEvolutionOfTheta_s}, we get: 

\begin{equation} \label{ContitionalTimeEvolutionOfTheta_s_Components}
\begin{aligned}
\left( U_\tau(\Theta_s)\right)_{\Bigl( \sum\limits_{\mu=1}^{m-1} s_\mu + j, \sum\limits_{\nu=1}^{n-1}s_\nu + k \Bigr)}
&=
\left(
\e^{-i \, J_m \, \tau} \, 
\hat{\bTheta}^{(m,n)} \, 
\e^{i \, J_n^T \, \tau} 
\right)_{j \, k} = \\
&=
\sum\limits_{\alpha=1}^{s_m} \,
\sum\limits_{\beta=1}^{s_n} \, 
\underbrace{
\left(\e^{-i \, J_m \, t} \right)_{j \, \alpha}
}_{
\e^{-i \, \delta_m \, t} \, \frac{t^{(\alpha-j)}}{(\alpha-j)!} \, 1_{\{j \leq \alpha\}}
} \, 
\left( \hat{\bTheta}^{m,n}\right)_{\alpha \, \beta} \, 
\underbrace{
\left(\e^{i \, J_m^T \, t} \right)_{\beta \, k}
}_{
\e^{i \, \delta_n \, t} \, \frac{t^{(\beta-k)}}{(\beta-k)!} \, 1_{\{k \leq \beta\}}
} = \\
&=
\sum\limits_{\alpha=j}^{s_m} \,
\sum\limits_{\beta=k}^{s_n} \,
\underbrace{
\e^{i \, (\delta_n^* - \delta_m) \, t}
}_{
\e^{i \, (R_n - R_m) \, t} \, \e^{-(I_n + I_m) \, t}
} \, 
\frac{t^{(\alpha-j)}}{(\alpha-j)!} \, 
\frac{t^{(\beta-k)}}{(\beta-k)!} 
\left(\hat{\bTheta}^{(m,n)} \right)_{\alpha, \beta}. 
\end{aligned}
\end{equation}


\subsubsection{Auxiliary Calculations for equation \eqref{ExpressionOf_I_s_of_tau_s} } \label{AuxiliaryCalculations_For_I_s_of_tau_s}

When $\Theta_s \in \Omega_{\brho_0}$ is a state in the quantum trajectory that is not a possible trapping state, then 

\begin{equation}\label{Auxiliary_I_s_of_tau_s_s}
\begin{aligned}
\lim\limits_{\tau \to \infty}
 \Tr[U_\tau(\Theta_s)] 
 \, \cdot \, 
 \int_0^\tau
 \frac{U_t(\Theta_s)}{\Tr[U_t(\Theta_s)]}
 \d \, t = 0. 
\end{aligned}
\end{equation}

\begin{proof}
It suffices to show that the left hand side of equation \eqref{Auxiliary_I_s_of_tau_s_s} is a positive semi-definite matrix with vanishing trace. To see this, we take an arbitrary vector $\psi\in \C^N$ and compute 

\begin{equation}
\begin{aligned}
\ScalProd{\psi}{U_t(\Theta_s) \, \psi}
=
\ScalProd{\e^{i \, H_c^\dagger \, t} \, \psi}{\Theta_s \, \e^{i \, H_c^\dagger \, t} \, \psi}
\geq \, 0, 
\end{aligned}
\end{equation}

where we used the fact that the density matrix $\Theta_s$ is positive semi-definite. Since multiplying by a positive scalar and integrating over the interval $[0,\tau]$ does not change positivity, we have shown the first part of our claim. 

For the second part we use the Jordan normal form of $H_c$ (see equation \eqref{JordanNormalFormOf_Hc})

We recall that for non-possible trapping states $\Theta_s$ the trace $U_t(\Theta_s)$ vanishes as $t$ tends to infinity (see the algorithm for the unravelling in section \ref{AlgorithmForunravelling}, 2 ii)) and compute

\begin{equation}\label{NotATrappingState}
\begin{aligned}
0 
&\xlongequal[\text{trapping state}]{\Theta_s\text{ is no possible }}
\lim\limits_{t \to \infty}
\Tr[U_t(\Theta_s)]
=
\lim\limits_{t \to \infty}
\sum\limits_{m=1}^M \, 
\sum\limits_{j=1}^{s_m } \, 
\underbrace{
\left( U_\tau(\Theta_s)\right)_{\Bigl( \sum\limits_{\mu=1}^{m-1} s_\mu+ j, \sum\limits_{\mu=1}^{m-1}s_\mu  + j \Bigr)}
}_{
\left(
\e^{-i \, J_m \, \tau} \, 
\hat{\bTheta}^{(m,m)} \, 
\e^{i \, J_m^T \, \tau} 
\right)_{j \, j} 
}=  \\
&=
\lim\limits_{t \to \infty}
\sum\limits_{m=1}^M \, 
\sum\limits_{j=1}^{s_m } \, 
\sum\limits_{\alpha, \beta=j}^{s_m} \,
\e^{-2 \, I_m \, t} \, 
\frac{t^{(\alpha-j)}}{(\alpha-j)!} \, 
\frac{t^{(\beta-j)}}{(\beta-j)!} 
\left(\hat{\bTheta}^{(m,m)} \right)_{\alpha, \beta}
\end{aligned}
\end{equation}

This means that whenever the imaginary part of an eigenvalue of the conditional Hamiltonian equals zero, then the corresponding block of $\Theta_s$ must vanish

\begin{equation}\label{Im_delta_k_0_implies_Theta_kk_0}
\begin{aligned}
\Im[\delta_m] = 0 \Longrightarrow \hat{\bTheta}^{(m,m)} = \bzero^{s_m \times s_m}. 
\end{aligned}
\end{equation}

When we now take the trace over the left side of equation \eqref{Auxiliary_I_s_of_tau_s_s}, we get 

\begin{equation}
\begin{aligned}
\lim\limits_{\tau \to \infty}
\Tr[U_\tau(\Theta_s)] 
\, \cdot \, 
\underbrace{
\Tr\left[
\int_0^\tau
\frac{U_t(\Theta_s)}{\Tr[U_t(\Theta_s)]}
\d \, t \right]
}_{
\tau
} 
\xlongequal{\eqref{Im_delta_k_0_implies_Theta_kk_0}}
0, 
\end{aligned}
\end{equation}

since multiplying equation \eqref{NotATrappingState} by a factor of $\tau$ does not influence the limiting behaviour. 
\end{proof}


\subsection{In a classical, discrete-time Markov chain with a finite state space a trajectory $(\Theta_n)_{n \in \N}$ will eventually be captured by a minimal absorbing set, that is $\Prob(\Theta_n \in \{\text{minimal absorbing sets}\}) = 1$. }\label{TrajectoriesAreCapturedByMinimalAbsorbingSets}

\begin{proof}
Let $B_1, \dots, B_n$ be the minimal absorbing sets of the associated state transition network for the Markov chain. We know that for every state $s \in \Omega$ in the network there is a path to a minimal absorbing set $B \subset \Omega$. This means that there is a finite number of states $s = s_0, \dots, s_L \in B$ with $q_{s_j \to s_{j+1}}>0$ \cite{fernengel2022obtaining}. We now let $l_i$ be the length of the shortest (non-zero) path to the minimal absorbing set number $i \in \{1, \dots, n\}$ and $q_i$ be the largest probability for reaching the set $B_i$ in $l_i$ time steps, namely

\begin{equation*}
\begin{aligned}
l_i
&:=
\min\limits_{l \in \N_0}
\left\{
(\omega_0 = s, \dots, \omega_l \in B_i) \,:\, q_{\omega_j \to \omega_{j+1}}>0 \text{ for all }j \in \{0, \dots, l\}
\right\} 
&&L
:=
\max\limits_{i \in \{1, \dots, n\} } \{l_i\} \\ 
\\
q_i 
&:=
\max\limits_{\boldsymbol{\omega} \in \Omega^{l_i+1}}
\left\{
\prod\limits_{j=0}^{l_i-1} q_{\omega_j \to \omega_{j+1}}, 
 \text{ with } \omega_0 = s \text{ and } \omega_{l_i} \in B_i
\right\}
&&q_0
:=
\min\limits_{i \in \{1, \dots, n\} } \{q_i\}. 
\end{aligned}
\end{equation*}

Further, be let $L \, (q_0)$ be the largest (smallest) of the $l_i \, (q_i)$. For a fixed natural number $m \in \N$ we define the event that a trajectory $\omega \in \Omega^\N$ after $L \cdot m$ time steps is not contained in a minimal absorbing set 

\begin{equation}
\begin{aligned}
A_m
&:=
\{
\omega \in \Omega^\N \,:\, 
\omega_{L \cdot m} \nin \bigcup\limits_{i=1}^n B_i
\}. 
\end{aligned}
\end{equation}

The corresponding probability for $A_m$ to occur can be estimated to $\Prob(A_m) \leq (1-q_0)^m$. Since the sum over all probabilities of $A_m$ is finite $(\sum\limits_{m \in \N} \Prob(A_m) < \infty)$, we know by the lemma of Borel-Cantelli that the probability for $A_m$ to occur infinitely often vanishes, $\Prob(A_m  \text{  infinitely often }) = 0$.

\end{proof}

\subsection{The time average of a classical, discrete-time Markov chain with finite state space}\label{TimeAverageOfADiscreteTimeMarkovChain}

Let $Q$ be the transition matrix of a discrete-time Markov chain that is the entries represent the transition probabilities $Q_{ij} = \Prob(j \to i)$ and $\bq_k(\bq_0) = Q^k \, \bq_0$ is the probability distribution after $k\in \N_0$ jumps. While the limit $k\to \infty$ need not exist, since oscillation are possible (consider the aperiodic Markov chain with the transition matrix $Q = \begin{pmatrix}
0 & 1 \\
1 & 0
\end{pmatrix}$), the time average does and equals

\begin{equation}
\begin{aligned}
\lim\limits_{K \to \infty} 
\frac{1}{K}
\sum\limits_{k=0}^{K-1}
Q^k \, \bq_0
=
\sum\limits_{i=1}^n \Prob(B_i \,|\, \bq_0) \, \bq^\infty(\bq_0 \in B_i), 
\end{aligned}
\end{equation}
where $B_1, \dotsc, B_n$ are the minimal absorbing sets of the associated state-transition network.

\begin{proof}
If $Q = S^{-1} \, J \,S $ is the Jordan normal form of the transition matrix, it can be shown by induction that the absolute value of the $i$-$j$-th entry of the $K$-th exponent of $J$ is proportional to the $K$-th exponent of the absolute value of the largest eigenvalue of $Q$, that is 

\begin{equation}
\begin{aligned}
\left| (J^k)_{ij} \right| 
\leq
C_{ij} \, \cdot \, |\lambda_S|^{k + i - j} \, \cdot \, 1_{\{i \leq j \leq i + k\}}, 
\end{aligned}
\end{equation}
where $C_{ij}$ is a constant that only depends on $i$ and $j$ and $\lambda_S := \arg\max\{|\lambda| \,:\, \lambda \in \sigma(Q)\}$. 

When $|\lambda_S|\leq 1$, $\lambda_S \neq 1$, we have 

\begin{equation}
\begin{aligned}
\left |
\left(
\frac{1}{K}
\sum\limits_{k=0}^{K-1}
J^k \,
\right)_{i\, j}
\right|
&\leq 
\frac{1}{K}
\sum\limits_{k=0}^{K-1}
\underbrace{
\left|
\, \left(
J^k 
\right)_{i \, j} \,  \, 
\right|
}_{
\leq \, C_{ij} \, \lambda_S^{k+i-j}
}
\leq 
C_{ij} \, \lambda_S^{i-j} \, 
\frac{1}{K}
\underbrace{
\left(
\sum\limits_{k=0}^{K-1} \lambda_S^k
\right)
}_{
\frac{1-\lambda_S^K}{1-\lambda_S}
}
&\xlongrightarrow{K \to \infty}
0. 
\end{aligned}
\end{equation}

Since we can interpret the matrix $\G_Q := Q - \Id$ as a transition matrix of a continuous-time Markov chain, and $\Kern(\G_Q - \lambda \, \Id) = \Kern(Q - (\lambda+1) \, \Id)$ we conclude that for the eigenvalue $\lambda=1$ of $Q$ the algebraic multiplicity $a_{\lambda=1}$ must coincide with the geometric multiplicity $d_{\lambda=1}$

\begin{equation}
\begin{aligned}
d_{\lambda=1}(Q)
=
d_{\lambda=0}(\G_Q = Q - \Id)
\xlongequal{\cite{fernengel2022obtaining}}
a_{\lambda=0}(\G_Q = Q - \Id)
=
a_{\lambda=1}(Q). 
\end{aligned}
\end{equation}

This means that we can write the transition matrix as $Q = S^{-1} \, 
\begin{pmatrix}
\Id & 0 \\
0 & \tilde{J}
\end{pmatrix}
\,S$, with $\lim\limits_{K \to \infty} \frac{1}{K} \sum\limits_{k=0}^{K-1} \tilde{J}^k = 0$. 

When writing the initial state as a linear combination of generalized eigenvectors $\bq_0 = \sum\limits_{i=1}^{|\Omega|} \mu_i \, \bh_{\lambda}^{(i)}(Q)$, we can compute the time average of the Markov chain as 

\begin{equation}
\begin{aligned}
\bq^\infty(\bq_0)
&:=
\lim\limits_{K \to \infty} \frac{1}{K} \sum\limits_{k=0}^{K-1} Q^k \, \bq_0
=
S^{-1}
\, 
\lim\limits_{K \to \infty} \frac{1}{K} \sum\limits_{k=0}^{K-1} \underbrace{
J^k
}_{
\begin{pmatrix}
\Id & 0 \\
0 & \tilde{J}^k
\end{pmatrix}
}\, S \;  \bq_0
=
S^{-1} \, 
\begin{pmatrix}
\Id & 0 \\
0 & 0
\end{pmatrix} \, 
S \, 
\underbrace{
\hspace*{3mm}
\bq_0
\hspace*{3mm}
}_{
\sum\limits_{i=1}^{|\Omega|} \mu_i \,  \bh_{\lambda}^{(i)}(Q)
}= \\
&=
\sum\limits_{i=1}^{|\Omega|}
\mu_i \, S \, 
\begin{pmatrix}
\Id & 0 \\
0 & 0
\end{pmatrix} \, 
\underbrace{
S^{-1} \, \bh_{\lambda}^{(i)}
}_{
1_{\{1, \dots, n\}(i) \, \be_i}
}
=
\sum\limits_{i=1}^n
\underbrace{
\hspace*{2mm}
\mu_i
\hspace*{2mm}
}_{
P(B_i\,|\, \bq_0)
}
\;\;
\underbrace{
\hspace*{2mm}
\bh_{\lambda=1}^{(i)}
\hspace*{2mm}
}_{
\bq^\infty(\bq_0 \in B_i)
}
\xlongequal{(*)}
\sum\limits_{i=1}^n 
P(B_i\,|\, \bq_0) \; 
\bq^\infty(\bq_0 \in B_i). 
\end{aligned}
\end{equation}

\end{proof}

In the last step (*) we have used the fact that the probability $\Prob(B_i \,|\, \bq_0)$ for a single trajectory to land in the minimal absorbing set $B_i$, provided it started with the initial distribution $\bq_0$, is given by the coefficient $\mu_i$ that is determined by the initial condition $\left(\bq_0 = \sum\limits_{i=1}^{|\Omega|} \mu_i \, \bh_{\lambda_i}^{(i)}\right)$ and the form of the stationary states $\bq^\infty(\bq_0 \in B_i) = \bh_{\lambda=1}^{(i)}$, for $i \in \{1, \dots, n\}$: 

\begin{equation}
\begin{aligned}
\Prob(B_i \,|\, \bq_0)
\xlongequal{\text{Def}}
\sum\limits_{j \in B_i}
(\bq^\infty(\bq_0))_j
=
\mu_i \, 
\underbrace{
\left(\sum\limits_{j \in B_i} \bq^\infty(\bq_0 \in B_i)\right)
}_{
1
}
=
\mu_i. 
\end{aligned}
\end{equation}

It can easily be verified that the coefficients $\mu_i$ form indeed a probability distribution, namely they are non-negative and their sum equals one. \\
Alternatively, we can interpret $\Prob(B_i \, |\, \bq_0)$ not only as the probability for a single trajectory being captured by the minimal absorbing set $B_i$, but also as the fraction of trajectories being captured by $B_i$, whose initial distribution is given by $\bq_0$. To be more concrete, let $\omega \in \Omega^\N$ be a trajectory starting at some state $\omega_0 =j \in \Omega$. We know from appendix \ref{TrajectoriesAreCapturedByMinimalAbsorbingSets} that such a trajectory will almost surely be captured by some minimal absorbing set $B\subset \Omega$, which means there exists a natural number $l_\omega \in \N_0$ such that after $l_\omega-1$ time steps the trajectory will only assume values in $B$ $(\omega_{k} \in B \text{ for all } k\geq l_\omega)$. Then we can compute the probability for $\omega$ to 

\begin{equation}
\begin{aligned}
\Prob(\omega \,|\, \omega_0 = j)
=
\prod\limits_{i=0}^{l_\omega-1} q_{\omega_i \to \omega_{i+1}}
=
\prod\limits_{i=0}^{l_\omega-1} Q_{ \omega_{i+1}, \omega_i}. 
\end{aligned}
\end{equation}

When we now set $\Unravel(B \,|\, \omega_0 = j )$ to be the set of all trajectories starting at some state $j\in \Omega$ that are eventually captured by the minimal absorbing set $B\subset \Omega$, $\Unravel(B \,|\, \omega_0 = j ):= \{\omega \in \Omega^\N \,:\, \omega_0 = j \text{ and }\omega_k \in B \text{ eventually }\}$. The probabilities of this set of trajectories can also be computed to

\begin{equation}
\begin{aligned}
\Prob(\Unravel(B) \,|\, \omega_0 = j)
=
\sum\limits_{\omega \in \Unravel(B \,|\, \omega_0 = j ) } \Prob(\omega \,|\, \omega_0 = j). 
\end{aligned}
\end{equation}

When we do not have a fixed stating point $j \in \Omega$, but an ensemble of starting points $\bq_0 = \sum\limits_{j=1}^{|\Omega|} \tilde{\mu}_j \, \be_j$, the corresponding probability equals

\begin{equation}
\begin{aligned}
\Prob(B \,|\, \bq_0)
:=
\Prob(\Unravel(B) \,|\, \bq_0)
= \sum\limits_{j=1}^{|\Omega|}\tilde{\mu}_j \, \Prob(B \,|\, \be_j)
=
\sum\limits_{j=1}^{|\Omega|} \,
\tilde{\mu}_j  \,
\sum\limits_{\omega \in \Unravel(B) } \,
 \Prob(\omega \,|\, \omega_0 = j). 
\end{aligned}
\end{equation}

So the conditional probability $\Prob(B \,|\, \bq_0)$ equals the sum over the probabilities over all trajectories, which land eventually in the minimal absorbing set $B\subset \Omega$ and whose initial conditions were distributed in $\Omega$ according to $\bq_0 \in (\R_{\geq \, 0})^{|\Omega|}$, with $\tilde{\mu}_j\geq \, 0$ being the fraction of trajectories that start at the state $j$ (note that we do not distinguish between the state $j \in \Omega$ and the corresponding unit vector $\be_j =(\underbrace{0, \dots, 0}_{j-1}, 1,\underbrace{0, \dots, 0}_{|\Omega|-j})^T$).

\subsection{The form of the (unique) stationary solution of a continuous-time Markov chain for a minimal absorbing set} \label{FormOfStationarySolutions}

\begin{figure}[H]
\begin{center}
\includegraphics[width=0.3\columnwidth]{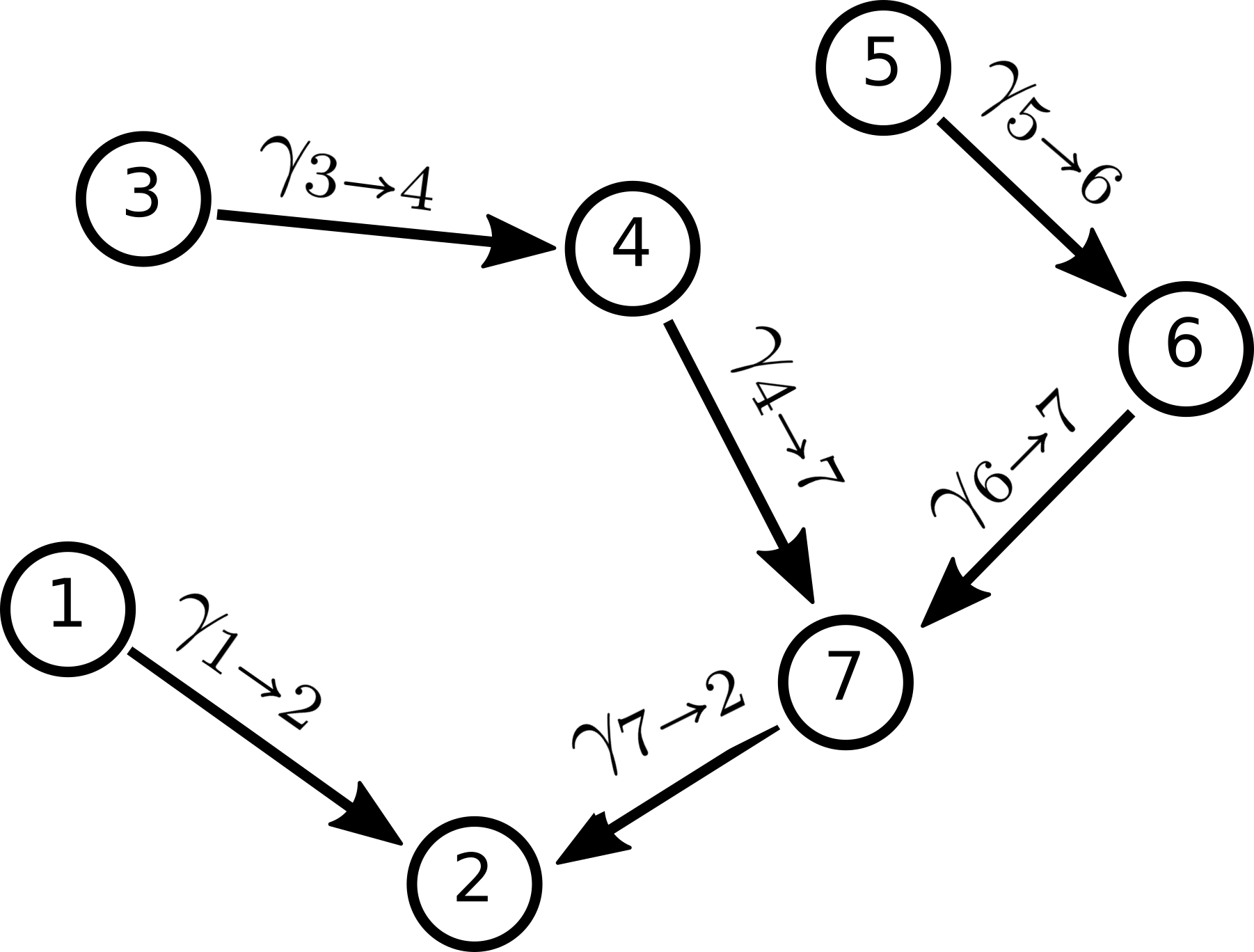} 
\caption{ Example of an in-tree rooted at state number $2$. There is a unique directed path from each state of the tree leading to state number $2$. Its weight equals $\prod\limits_{(i,j) \in \Edge(\Omega)} \, \g_{i\to j} = \, \g_{1\to 2} \, \g_{7\to 2} \, \g_{7\to 2} \, \g_{4 \to 7} \, \g_{6 \to 7} \, \g_{3 \to 4} \, \g_{5 \to 6}$}
\label{Example_InTreeRootedAtState_2} 
\end{center}
\end{figure}

We call a network $\Omega$ with no self-loops an \textbf{in-tree} (also called \textbf{anti-arborescence} \cite{gabow1978finding}) \textbf{rooted at state $\omega_0 \in \Omega$} if for all states $\omega \in \Omega$ there is a unique directed path leading from state $\omega$ towards the root $\omega_0$. 
An example is given in Figure \ref{Example_InTreeRootedAtState_2}, where the root is state number $2$. 
The \textit{weight} of an in-tree is the product of the link-strength of all edges of the in-tree: $\prod\limits_{(i,j) \in \Edge(\Omega)} \, \g_{i\to j}$ 
, with $\Edge(\Omega)$ denoting the set of edges of the network. 

When $\G$ is generator for the so-called \textit{Master equation} $\dot{\bp} = \G \, \bp$, with $\G_{ij} = 
\begin{cases}
 \g_{j\to i}  & ,  i \neq j \\
  -\mathlarger{\sum}\limits_{k=1}^{|\Omega|} \g_{j\to k} &  , i = j\, .
\end{cases}$,

it can be shown (\cite{chaiken1978matrix, mirzaev2013laplacian, fernengel2022obtaining}) that the unique stationary solution of the Master equation for a strongly connected network with no self-loops can be computed according to equation \eqref{FormulaFor_p_infty}, where $\InTree_{i \, \leftarrow}(\Omega)$ is the set of all in-trees of $\Omega$, which are rooted in state number $i \in \Omega$. An example is given in figure \eqref{Example_StationaryState_InTrees} below.

\begin{equation}\label{FormulaFor_p_infty}
\bp_\infty
=
\colvec{3}{
\mathlarger{\mathlarger{\sum\limits}}_{T\, \in \, \InTree_{1 \, \leftarrow}} \; \mathlarger{\mathlarger{\prod\limits}}_{(i,j) \, \in \,  \Edge(T) } \g_{i \to j} 
}{\\ \vdots \\ }{
\mathlarger{\mathlarger{\sum\limits}}_{T\, \in \, \InTree_{|\Omega| \, \leftarrow}} \; \mathlarger{\mathlarger{\prod\limits}}_{(i,j) \, \in \,  \Edge(T)  } \g_{i \to j} 
}.
\end{equation}

\begin{figure}[H]
\begin{minipage}{0.3\textwidth}
\begin{center}
\begin{subfigure}{1.0\textwidth}
\subcaption{}
\label{Example_StationaryState_OriginalNetwork}
\includegraphics[width=0.8\columnwidth]{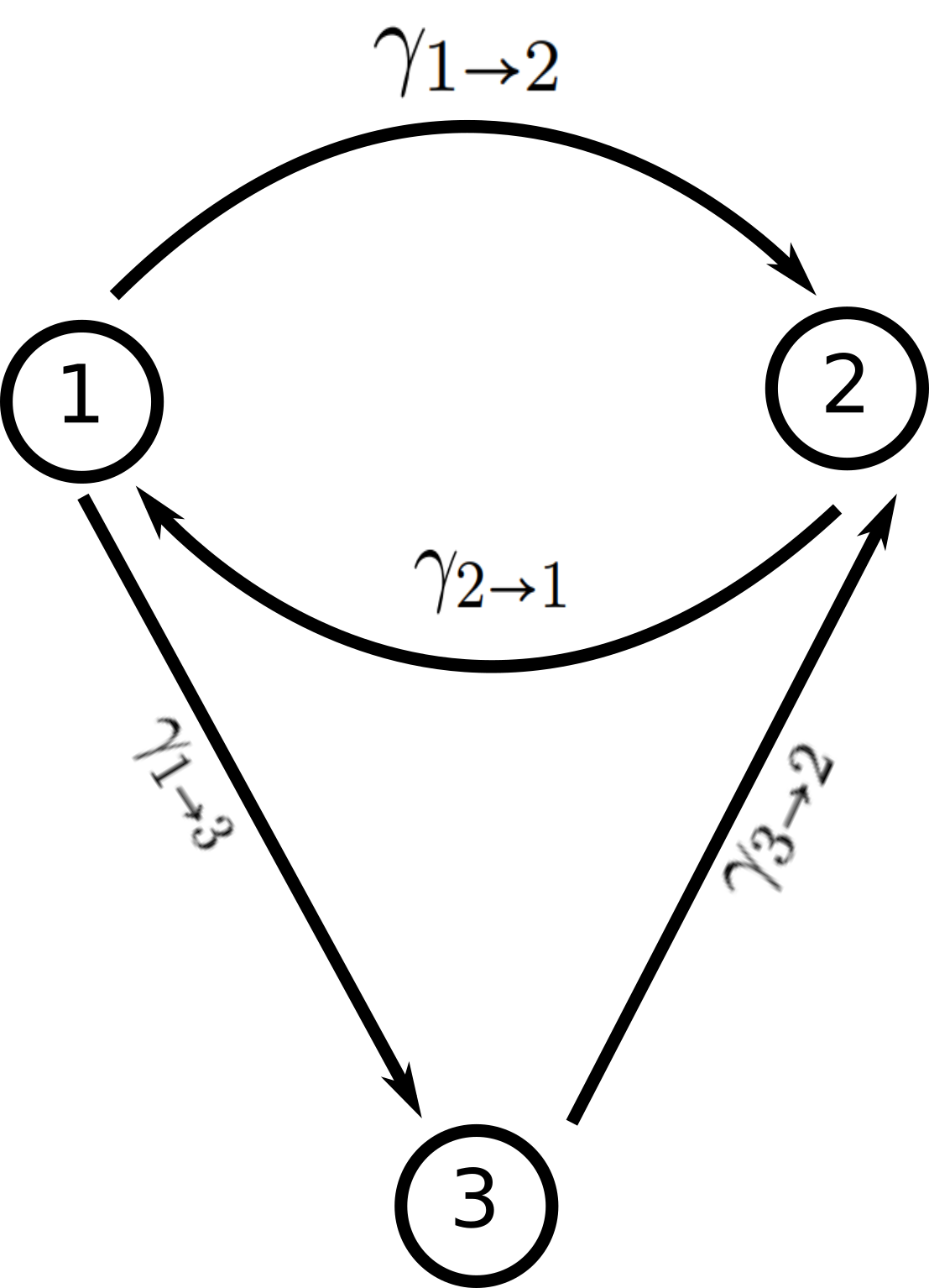}   
\end{subfigure}
\end{center}
\end{minipage} 
\hspace*{3mm} 
\begin{minipage}{0.7\textwidth}
\begin{center}
 \begin{subfigure}{0.25\textwidth}
\subcaption{In-tree rooted in $1$}
\label{Example_StationaryState_InTrees_LeadingTo_1}
\includegraphics[width=1.0\columnwidth]{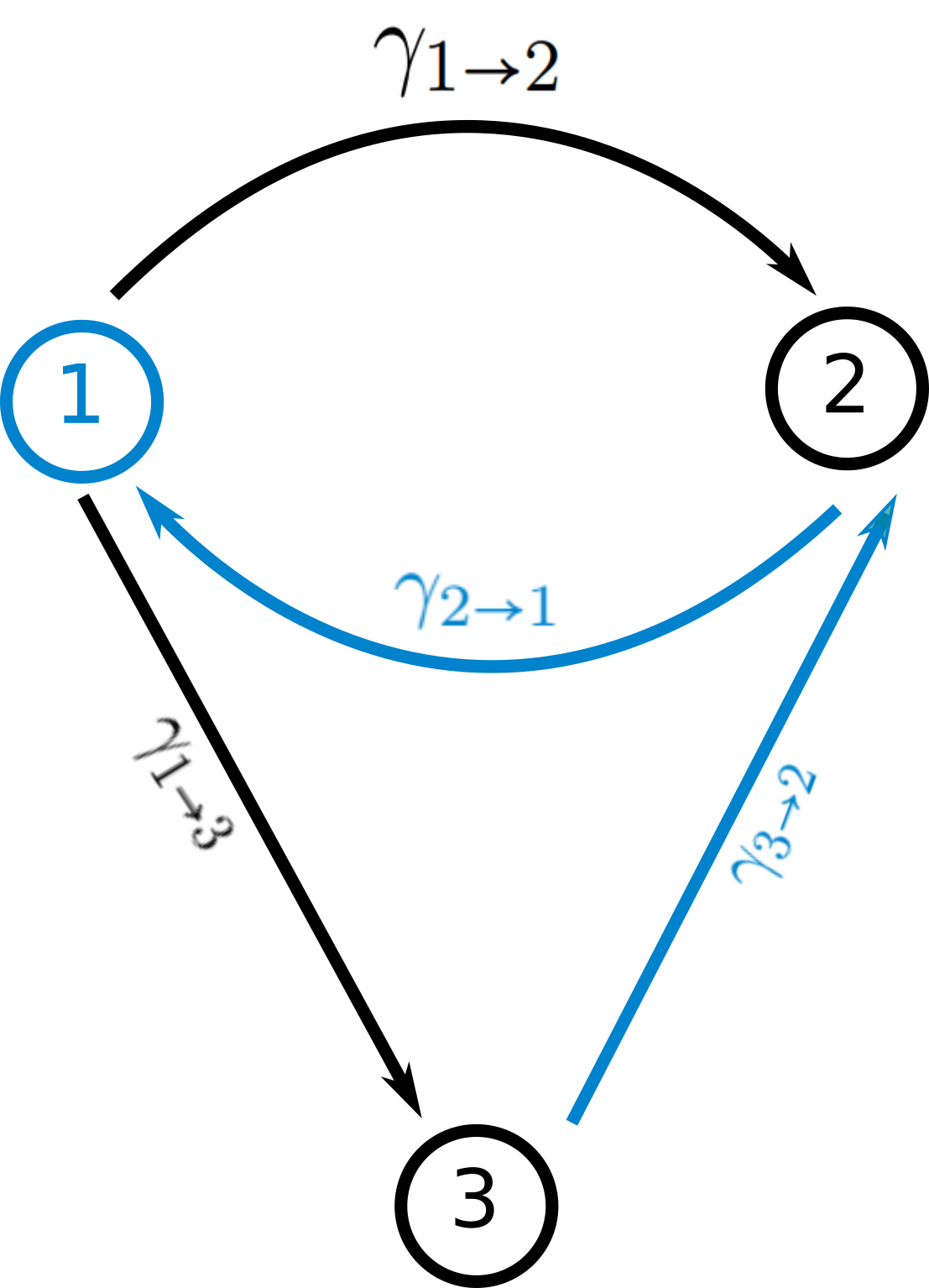} 
\end{subfigure} \hspace*{3mm} \begin{subfigure}{0.25\textwidth}
\subcaption{In-tree rooted in $3$}
\label{Example_StationaryState_InTrees_LeadingTo_3}
\includegraphics[width=1.0\columnwidth]{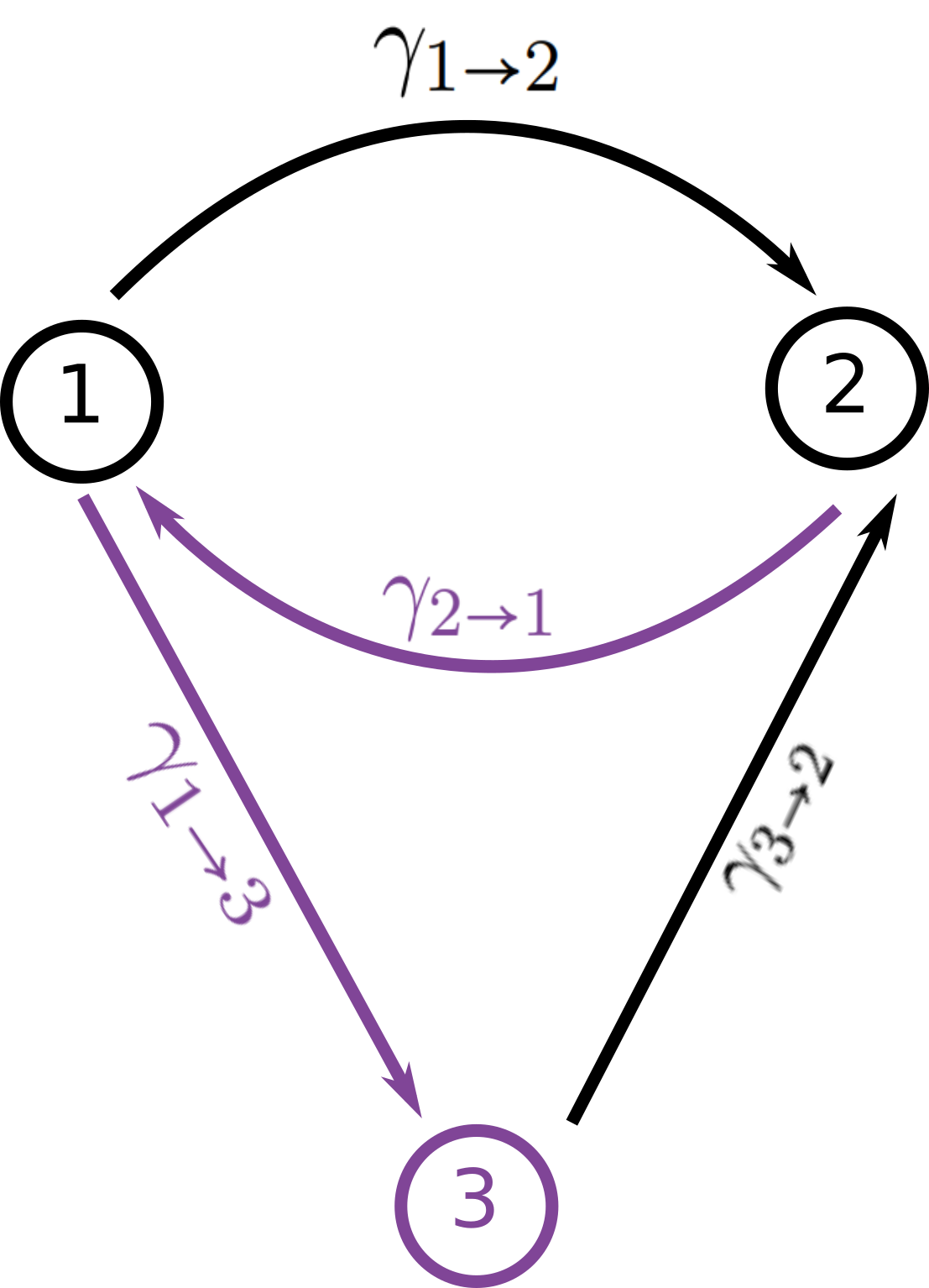} 
\end{subfigure} \\
\vspace*{7mm}
 \begin{subfigure}{0.25\textwidth}
\subcaption{First in-tree rooted in $2$}
\label{Example_StationaryState_InTrees_LeadingTo_2a}
 \includegraphics[width=1.0\columnwidth]{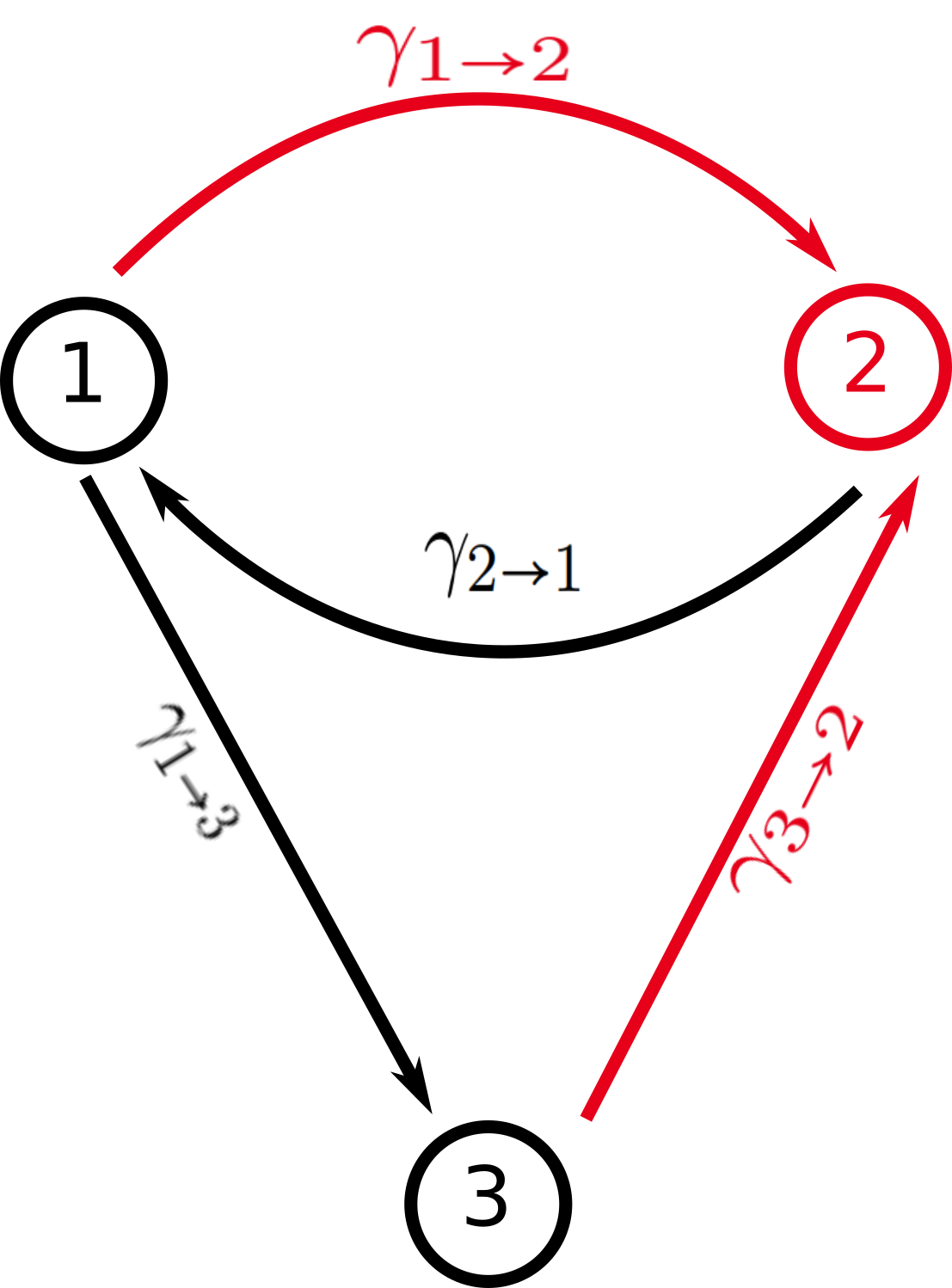}
\end{subfigure} \hspace*{3mm} \begin{subfigure}{0.25\textwidth}
\subcaption{Second in-tree rooted in $2$}
\label{Example_StationaryState_InTrees_LeadingTo_2b}
 \includegraphics[width=1.0\columnwidth]{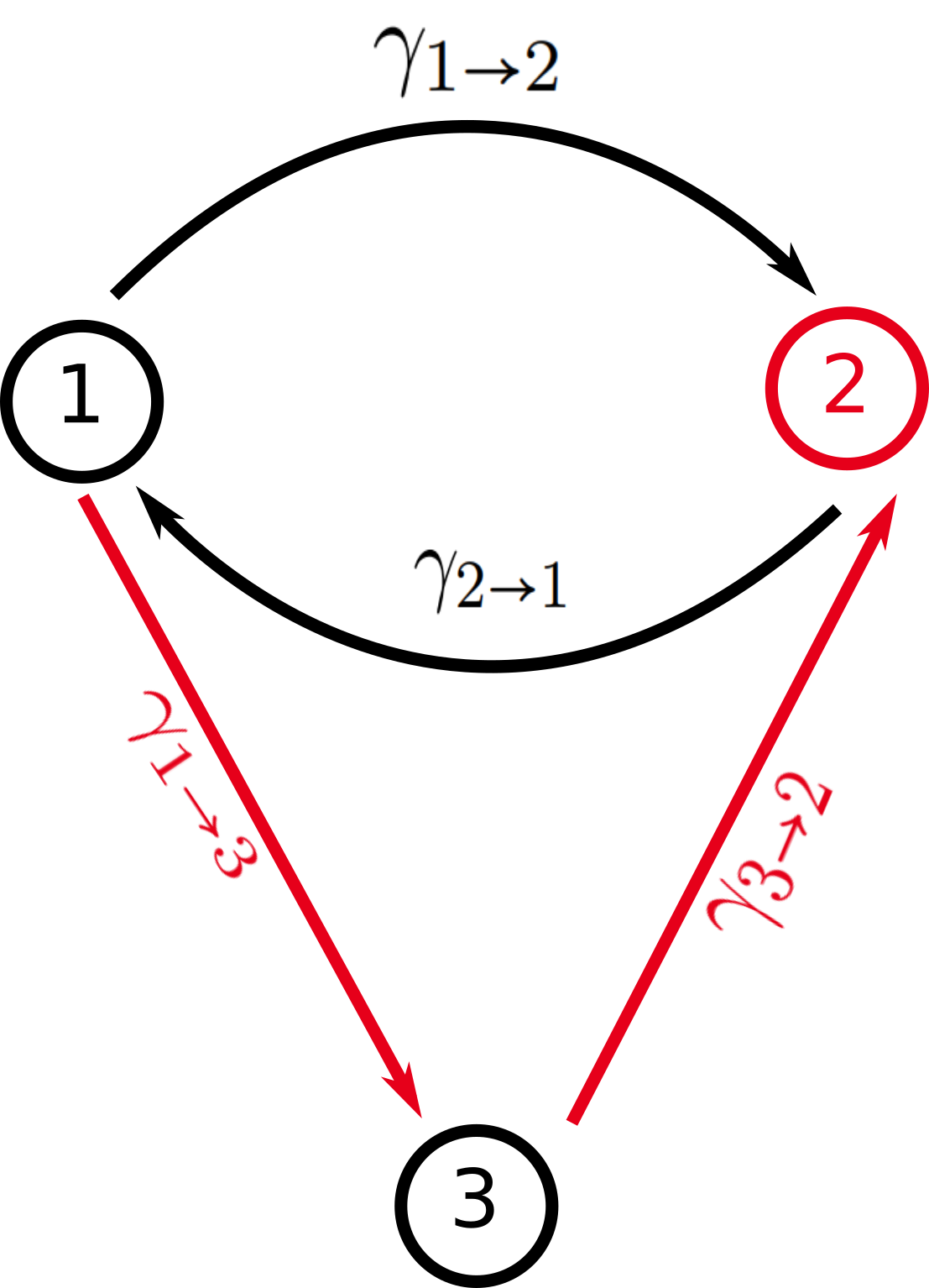} 
 \end{subfigure}
 \end{center}
\end{minipage}
\caption{ Example of a strongly connected network (Figure \ref{Example_StationaryState_OriginalNetwork}) together with the corresponding in-trees rooted in states number { \color{TUDa_2b} $1$ (Figure \ref{Example_StationaryState_InTrees_LeadingTo_1}) } and { \color{TUDa_11a} $3$ (Figure~\ref{Example_StationaryState_InTrees_LeadingTo_3})},  and the two in-trees rooted in state  number {\color{TUDa_9b} $2$ (Figure~\ref{Example_StationaryState_InTrees_LeadingTo_2a} and \ref{Example_StationaryState_InTrees_LeadingTo_2b})}. The kernel of $\G$ is the span of a vector whose $i$-th component is the sum over all in-trees rooted in state number $i$ of the product of the rates of all edges that constitute that particular in-tree. In this example we have \\
\texorpdfstring{$ \Kern(\G) = \Span \left\{ \colvec{3}{ \color{TUDa_2b} \g_{3 \to 2} \cdot \g_{2 \to 1} }{ \color{TUDa_9b} \g_{1 \to 2} \cdot \g_{3 \to 2} + \g_{1 \to 3} \cdot \g_{3 \to 2}   }{\color{TUDa_11a} \g_{2 \to 1} \cdot \g_{1 \to 3}} \right\}$}{Kern}. }
\label{Example_StationaryState_InTrees}
\end{figure}

For the discrete-time Markov associated to a strongly connected, weighted network with possible self-loops, we know from appendix \ref{TimeAverageOfADiscreteTimeMarkovChain} that the stationary solution 

\begin{equation}\label{q_infty_for_StronglyConnected}
\begin{aligned}
\bq^\infty
:=
\lim\limits_{K\to \infty} \, \frac{1}{K} \, 
\sum\limits_{k=0}^{K-1} \, Q^k \, \bq_0
\in
\Kern(Q - 1 \cdot \Id)
\end{aligned}
\end{equation}

exists and is independent for all initial condition $\bq_0 \in (\R_{\geq \, 0})^{|\Omega|}$. 

Since we can interpret $\G_Q := Q - \Id$ as the generator for a continuous time Markov chain, we can compute the compute the expression in equation \eqref{q_infty_for_StronglyConnected} to

\begin{equation}\label{FormulaFor_q_infty}
\begin{aligned}
\bq^\infty
=
\colvec{3}{
\mathlarger{\mathlarger{\sum\limits}}_{T\, \in \, \InTree_{1 \, \leftarrow}} \; \mathlarger{\mathlarger{\prod\limits}}_{(i,j) \, \in \,  \Edge(T) \atop i\neq j} q_{i \to j} 
}{\\ \vdots \\ }{
\mathlarger{\mathlarger{\sum\limits}}_{T\, \in \, \InTree_{|\Omega| \, \leftarrow}} \; \mathlarger{\mathlarger{\prod\limits}}_{(i,j) \, \in \,  \Edge(T)  \atop i\neq j } q_{i \to j} 
}.
\end{aligned}
\end{equation}

\subsection{The need for a finite state space}\label{TheNeedForAFiniteStateSpace}
Now that we have derived a formula for the stationary solution of the Lindblad equation, which is valid when certain assumptions are satisfied, we will look at these assumptions and check whether they can be dropped. 

One crucial assumptions was the need for a finite state space $|\s|<\infty$. When dropping this assumption, the formula does no longer hold true in general, as the following counter-example shows:

\begin{equation}\label{CounterExample}
\begin{aligned}
V_1 = \begin{pmatrix}
 1 & 0 \\
 0 & \frac{1}{\sqrt{2}}
 \end{pmatrix}, \hspace*{3mm}
V_2 = \begin{pmatrix}
1/2 &  0 \\
1/2 & 0 
\end{pmatrix}, \hspace*{3mm}
\Lambda = \sum\limits_{k=1}^2 \g_k V_k^\dagger \, V_k
= \begin{pmatrix}
\g_1 & 0  \\
0 & \frac{\g_1 + \g_2}{2} 
\end{pmatrix}
\end{aligned}
\end{equation}

The set of possible states appearing in the Markov chain can be computed to

\begin{equation}
\begin{aligned}
\Omega_{\brho_0}
=
\Bigl\{ \brho_0, \,   \Theta_{s_{k}}\;:\: k \in \N_0 \Bigr\}
\end{aligned}
\end{equation}
with 

\begin{equation}
\begin{aligned}
\Theta_{s_{2 \, k}}
&=
\frac{1}{\frac{\brho_{22}(0)}{2^{k}} \,  + \brho_{11}(0) } \, 
\begin{pmatrix}
\brho_{11}(0) & \frac{\brho_{11}(0)}{2^{k/2}} \\
\frac{\brho_{21}(0)}{2^{k/2}} & \frac{\brho_{22}(0)}{2^{k}}   \\
\end{pmatrix}, \\ \\ 
\Theta_{s_{2 \, k+1}}
&=
\frac{1}{\frac{1}{2^{k}} \, +1 } \, 
\begin{pmatrix}
1 & \frac{1}{2^{k/2}} \\
\frac{1}{2^{k/2}} & \frac{1}{2^{k}}   \\
\end{pmatrix}. 
\end{aligned}
\end{equation}

The probability for applying the Lindblad operators $V_1$ and $V_2$ is given by

\begin{equation}
\begin{aligned}
\Prob(\pi_{k+1} =1 \,|\, \Theta_k)
=
1 - \frac{\g_2}{\g_1 + \g_2}
\, \cdot \, 
\frac{\left(\Theta_n\right)_{22}}{2^k \, \left(\Theta_n\right)_{11} + \left(\Theta_n\right)_{22}} \\
\Prob(\pi_{k+1} =2 \,|\, \Theta_k)
=
\frac{\g_2}{\g_1 + \g_2}
\, \cdot \, 
\frac{\left(\Theta_n\right)_{22}}{2^k \, \left(\Theta_n\right)_{11} + \left(\Theta_n\right)_{22}}. 
\end{aligned}
\end{equation}

Then, by the lemma of Borel-Cantelli, the probability for applying the operator $V_2$ infinitely often is zero (for $\brho_{11}(0) \neq 0$), since $\sum\limits_{k=0}^\infty \Prob(\pi_{k+1} =2 \,|\, \Theta_k) < \infty $. 

\begin{figure}[H]
\begin{center}
\includegraphics[width=0.8\columnwidth]{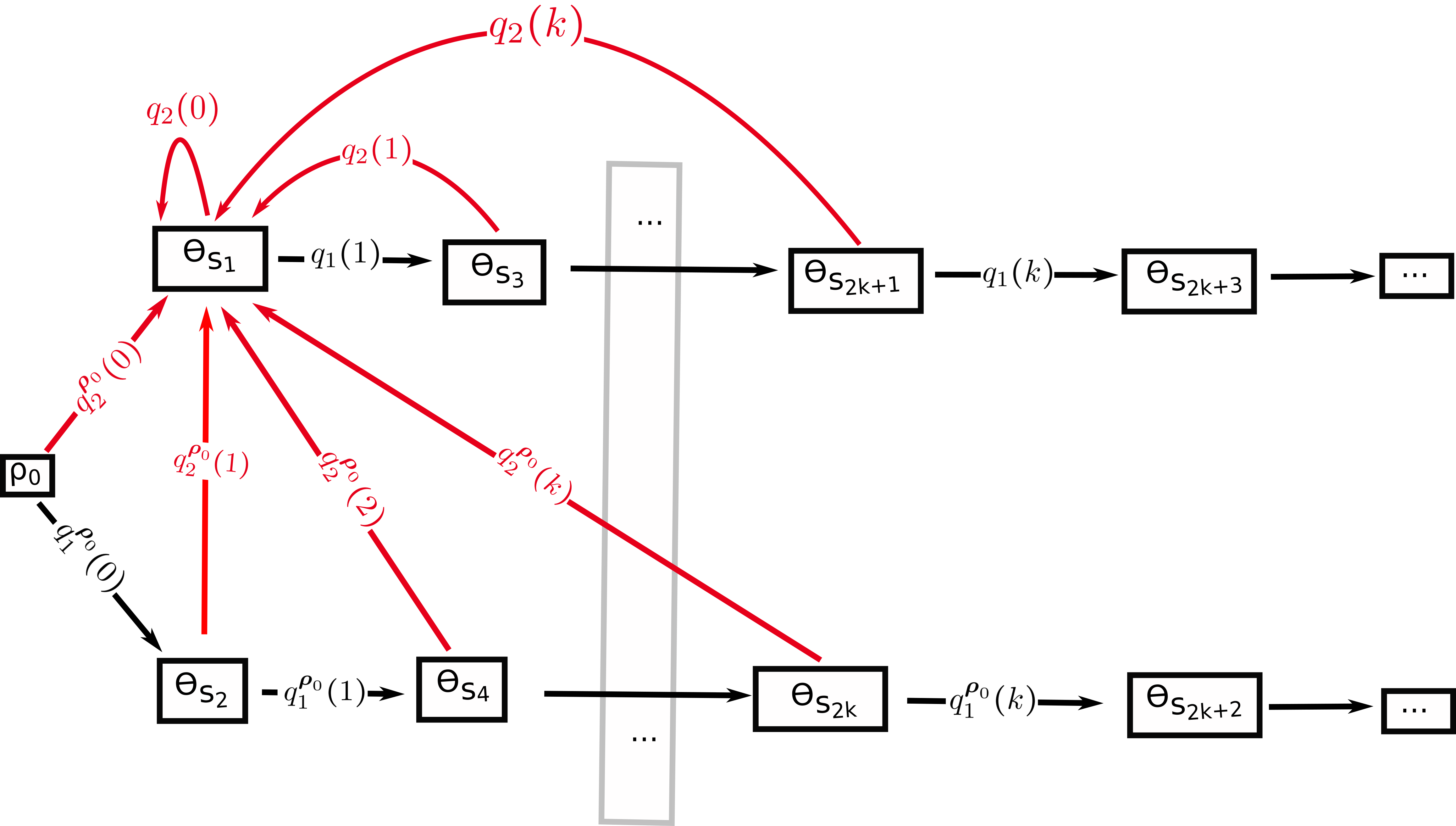} 
\caption{State transition network for the the Lindblad operators of example \ref{CounterExample}. All states are transient states and the system will eventually follow only the black transitions, indicating the application of operator $V_1$.  }
\label{StateTransitionNetwork_CounterExample_InfiniteNumberOfStates_2dim} 
\end{center}
\end{figure}

Hence we know almost surely that from a certain time step $k \in \N$ onward only the Lindblad operator $V_1$ will be applied (so the system will eventually follow the black arrow of figure \ref{StateTransitionNetwork_CounterExample_InfiniteNumberOfStates_2dim}). This means that all states are transient states and hence no stationary distribution for this Markov chain exists. 

\subsection{When the number of quantum jumps in a quantum trajectory is not bounded, then $\frac{t_{J(T)}}{T} \xlongrightarrow{T \to \infty}1$ in probability. }\label{t_J(T)_DividedBy_T_TendsTo_1}

\begin{proof}
Fix $\epsilon > 0$, then we have: 

\begin{equation}
\begin{aligned}
\Prob\left( 
\left|
\frac{t_{J(T)}}{T} - 1
\right| > \epsilon
\right)
=
\Prob \left(T - t_{J(T)} > \epsilon \, T\right)
=
\Tr[U_{\epsilon \, T} \,  (\Theta_{J(T)})]
\xlongrightarrow{T \to \infty} 0, 
\end{aligned}
\end{equation}

where we used the fact that $\Theta_{J(T)}$ can never be a possible trapping state, since the number of quantum jumps is by assumption not bounded, so the quantum trajectory will almost surely jump at some point.

\end{proof}
\subsection{Illustrating the algorithm for obtaining the stationary probabilities of a discrete-time Markov chain}\label{IllustratingTheAlgorithmForObtainingTheStationaryProbabilitiesOfADiscrete-timeMarkovChain}

\begin{figure}[H]
\centering
\includegraphics[width=0.25\columnwidth]{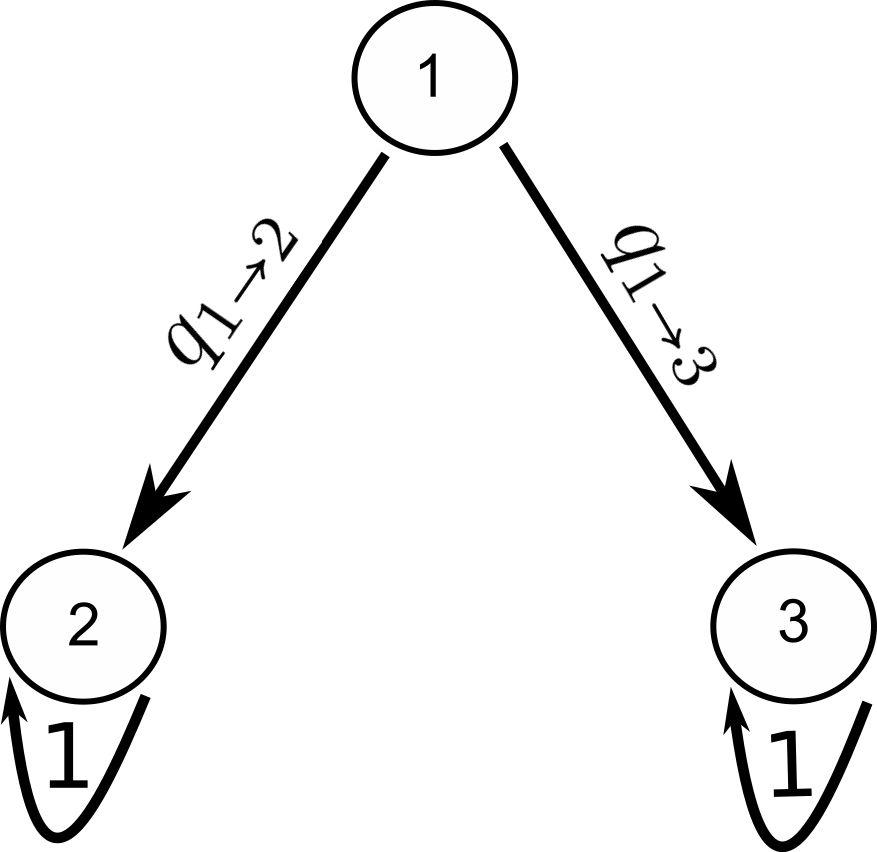}
\caption{Example of a state transition network of a Markov chain whose stationary probability distribution $\bq^\infty (\bq_0 )$ is given by equation \eqref{Example_StationaryProbability_MarkovChain}. }
\label{DemonstrationOfStationaryState}
\end{figure}

When the state transition network is given by figure \ref{DemonstrationOfStationaryState}, the initial distribution 

\begin{equation}
\begin{aligned}
\bq_0 
&=
\colvec{3}{\bq_1(0)}{\bq_2(0)}{\bq_3(0)} = \\
&=
\left( q_2(0) + q_1(0) \frac{q_{1\to 2}}{q_{1\to 2} + q_{1\to 3}} \right)\, \colvec{3}{0}{1}{0} + 
\left( q_3(0) + q_1(0) \frac{q_{1\to 3}}{q_{1\to 2} + q_{1\to 3}}  \right)\, \colvec{3}{0}{0}{1} + 
q_1(0) \colvec{3}{1}{-\frac{q_{1\to 2}}{q_{1\to 2} + q_{1\to 3}} \vspace*{1mm}}{-\frac{q_{1\to 3}}{q_{1\to 2} + q_{1\to 3}}}
\end{aligned}
\end{equation}
 is re-distributed after one time step to the following stationary distribution: 

\begin{equation}\label{Example_StationaryProbability_MarkovChain}
\begin{aligned}
\bq^\infty (\bq_0 )
&=
\underbrace{
\left( q_2(0) + q_1(0) \frac{q_{1\to 2}}{q_{1\to 2} + q_{1\to 3}} \right)
}_{
\Prob(B_1 = \{2\} \,|\, \bq_0)
}
\, 
\underbrace{
\colvec{3}{0}{1}{0}
}_{
\bq^\infty(\bq_0 \in \{2\})
}+ 
\underbrace{
\left( q_3(0) + q_1(0) \frac{q_{1\to 3}}{q_{1\to 2} + q_{1\to 3}}  \right)
}_{
\Prob(B_2 = \{3\}\,|\, \bq_0)
}\, 
\underbrace{
\colvec{3}{0}{0}{1}
}_{
\bq^\infty(\bq_0 \in \{3\}) 
} = \\ \\
&=
\Prob(B_1 = \{2\}\,|\, \bq_0) \;\bq^\infty(\bq_0 \in \{2\}) +
\Prob(B_2 = \{3\}\,|\, \bq_0) \; \bq^\infty(\bq_0 \in \{3\}). 
\end{aligned}
\end{equation}

\bibliographystyle{alpha}
\bibliography{literature} 

\end{document}